\documentclass{aa}

\newlength{\picwd}
\newlength{\thinpic}
\usepackage[utf8]{inputenc}
\usepackage{natbib}
\usepackage[above,below]{placeins}
\usepackage[dvipsnames]{xcolor}

\usepackage{gensymb}

\usepackage{txfonts}

\newcommand{\e}[1]{\ensuremath{\times 10^{#1}}} %
\newcommand{\un}[1]{\ensuremath{\ \mathrm{#1}}}

\newcommand{\rsun}{\ensuremath{\ R_\odot}}
\newcommand{\rot}[1]{\ensuremath{\nabla\times {#1}}}
\newcommand{\diver}[1]{\ensuremath{\nabla\cdot {#1}}}

\newcommand{\prop}[1]{}

\newcommand{\rev}[1]{#1}

\begin{document}

\title{Solar wind rotation rate and shear at coronal hole boundaries, possible consequences for magnetic field inversions.}

\authorrunning{R. F. Pinto, et. al}
\titlerunning{Solar wind speed and rotation shear in the corona}

\author{
  R. F. Pinto\inst{1,2},
  N. Poirier\inst{2},
  A. P. Rouillard\inst{2},
  A. Kouloumvakos\inst{2},
  L. Griton\inst{2},
  N. Fargette\inst{2},\\
  R. Kieokaew\inst{2},
  B. Lavraud\inst{2,3},
  A. S. Brun\inst{1}
}

\institute{
  Département d'Astrophysique/AIM, CEA/IRFU, CNRS/INSU, Univ. Paris-Saclay \& Univ. de Paris, 91191 Gif-sur-Yvette, France \\
\and 
  IRAP, Université de Toulouse; UPS-OMP, CNRS; 9 Av. colonel Roche, BP 44346, F-31028 Toulouse cedex 4, France\\
  \email{rui.pinto@irap.omp.eu} \\
\and
Laboratoire d'Astrophysique de Bordeaux, Univ. Bordeaux, CNRS, B18N, allée Geoffroy Saint-Hilaire, 33615 Pessac, France
}

\date{Received September 15, 1996; accepted March 16, 1997}

\abstract
{%
  In situ measurements by several spacecraft have revealed that the solar wind is frequently perturbed by transient structures that have been interpreted as magnetic folds, jets, waves and flux-ropes that propagate rapidly away from the Sun over a large range of heliocentric distances. 
  Parker Solar Probe (PSP), in particular, has detected very frequent rotations of the magnetic field vector at small heliocentric radial distances, accompanied by surprisingly large solar wind rotation rates.
  The physical origin of such magnetic field bends and switchbacks, the conditions for their survival across the interplanetary space, and their relation to solar wind rotation are yet to be clearly understood.
  }
  {
    \rev{We aim at characterising the global properties of the solar wind flows crossed by PSP,  at relating those flows to the rotational state of the low solar corona, and at identifying regions of the solar surface and corona that are likely to be sources of switchbacks and bends.}
  }
  {
    We traced measured solar wind flows from the spacecraft position down to the surface of the Sun to identify their potential source regions, and used a global magneto-hydrodynamic (MHD) model of the corona and solar wind to analyse the dynamical properties of those regions.
    We identify regions of the solar corona for which solar wind speed and rotational shear are important \rev{and long-lived, }that can be favourable to the development of magnetic deflections and to their propagation across extended heights in the solar wind.
    
  }
  {
    We show that coronal rotation is highly structured, and that enhanced flow shear and magnetic field  gradients develop near the boundaries between coronal holes and streamers\rev{, around and above pseudo-streamers, even when such boundaries are aligned with the direction of solar rotation}.
    The exact properties and amplitudes of the shears are a combined effect of the forcing exerted by the rotation of the corona and of its magnetic topology.
    A large fraction of the switchbacks identified by PSP map back to these regions, both in terms of instantaneous magnetic field connectivity and of the trajectories of wind streams that reach the spacecraft.
  }
  {
    \rev{We conclude that these regions of strong shears are likely to leave an imprint on the solar wind over large distances and to increase the transverse speed variability in the slow solar wind. The simulations and connectivity analysis suggest they can be a source of the switchbacks and spikes observed by Parker Solar Probe.}
    }


\maketitle

\section{Introduction}
\label{sec:intro}

Measurements made by Parker Solar Probe \citep[PSP;][]{fox_solar_2016} during its first set of orbits have revealed several intriguing properties of the solar wind at heliocentric distances that had never been probed before. 
PSP has shown that strong magnetic perturbations in the form of localised magnetic field-line bends, the most intense of which are termed switchbacks (SB), are omnipresent in the pristine solar wind measured near the Sun \citep{bale_highly_2019}.
PSP observations have also shown that the magnitude of the transverse (i.e, rotational) speeds of the solar wind increases rapidly in the corona with increasing proximity to the solar surface, up to amplitudes that were not expected considering its general radial trend higher up in the heliosphere \citep{kasper_alfvenic_2019}.
However, current knowledge of the exact way that the solar rotation propagates into and establishes in the highly magnetised solar corona is insufficient to interpret adequately the rotation state of the observed solar wind flows. 
Physical links between these coincidental phenomena are yet to be identified, but possibly relate to the structure of the solar wind and magnetic field at large scales, and to how their interplay generates regions with contrasting magnetic field directions and flows.
\rev{Interchange reconnection has been pointed out as a potential candidate mechanism \citep[][]{fisk_global_2020}, as it links large-scale rotation to magnetic reconnection.
  Interchange can occur at a variety of scales in the solar corona, from polar plumes above small magnetic bipoles inside coronal holes \citep{wang_footpoint_2004,wang_nature_2012,owens_heliospheric_2013} to streamer/coronal hole boundaries}.

Sudden reversals of the magnetic field have been observed for several decades.
Early examples can be spotted in observations from the Helios mission \citep{behannon_alfven_1981}.
These reversals have since been studied at various distances from the Sun and related to different photospheric, coronal or heliospheric phenomena \citep{kahler_topology_1996,ballegooijen_magnetic_1998,balogh_heliospheric_1999,yamauchi_differential_2004,yamauchi_magnetic_2004,velli_coronal_2011,neugebauer_evidence_2012,wang_footpoint_2004,wang_nature_2012,owens_heliospheric_2013,neugebauer_double-proton_2013,matteini_dependence_2014,borovsky_plasma_2016-1,horbury_short_2018,sterling_coronal-jet-producing_2020}.
SB's observed by PSP display a high degree of alfvenicity (cross-helicity), indicating that they propagate outwards mostly in the form of incompressible MHD waves \citep{bale_highly_2019}, that they correspond to a rotation of the magnetic field vector without change to its absolute value \citep{mozer_switchbacks_2020}, and that equipartition between the transverse velocity ($\delta v_\perp$) and magnetic field ($\delta b_\perp/\sqrt{\rho}$) perturbations is observed.
The physical origin and evolution of magnetic SBs across vast heliocentric distances remains elusive to this date. Currently debated hypothesis interpret them as products of surface and coronal dynamics (reconnection, jets, plumes) or, inversely, claim them to be formed in-situ in the heliosphere (turbulence, large-scale wind shear).
Numerical simulations of alfvénic perturbations by \citet{tenerani_magnetic_2020} suggest that switchbacks originating in the lower corona may survive out to PSP distances on their own, but only as long as they propagate across a sufficiently unperturbed background solar wind (free of significant density fluctuations, flow and magnetic shears that could destabilize them, e.g., the parametric instability).
\citet{owens_signatures_2020}, however, argue that solar wind speed shear is an essential factor for the survival of heliospheric magnetic field inversions (switchbacks) produced close to the Sun up to $1\un{AU}$, as these should not last long enough without being amplified by solar wind speed shear along their propagation path. 
\citet{macneil_evolution_2020} have furthermore shown that these magnetic inversions grow in amplitude and in frequency with altitude in HELIOS data, favouring the idea that they are either created or amplified by favourable wind shear in the heliophere.
From a different perspective, \citet{squire_-situ_2020} propose that magnetic SBs are a natural result of solar wind turbulence, and should therefore be produced in loco throughout the heliosphere.
\citet{ruffolo_shear-driven_2020} furthermore suggest that shear-driven MHD turbulence is capable of producing magnetic SBs, especially on the regions just above the Alfvén surface, where the observed solar wind flow density transitions from a striated to a flocculated pattern \citep[\emph{cf.}][]{deforest_fading_2016}.

Large-scale solar wind shear builds up naturally from two different components: gradients in wind speed in the meridional plane (transitions between slow and fast wind streams, between open and closed field regions), and gradients in azimuthal wind speed (rotation).
If the role of the former is reasonably easy to identify on large scales, that of the latter is much less well understood.
The solar photosphere is known to rotate with a well-defined differential latitudinal profile (with the equator having a higher rotation rate than the poles).
The corona above, albeit magnetically rooted in the photosphere, seems to exhibit a different rotation pattern, with regions that often appear to be rotating rigidly \citep{antonucci_rigid_1974,fisher_rotational_1984}.
It has been suggested that this rigidity could be the result of the interplay between emerging magnetic flux and the global field, involving sustained magnetic reconnection \citep{wang_quasi-rigid_1988,nash_mechanisms_1988}.
Overall, the coronal plasma seems to rotate with a more solid-body like pattern than the photosphere, and the more so at mid and high coronal altitudes  \citep{insley_differential_1995, bagashvili_statistical_2017}.
The rotation profile of the corona also seems to evolve in time, and to be linked to the solar cycle phase or, at least, to the specific coronal magnetic field configuration at a given moment \citep{badalyan_cyclic_2006,badalyan_two_2010}.
Observations by \citet{giordano_coronal_2008} using SoHO/UVCS have shown that a number of features superpose to the average large-scale latitudinal trends of the coronal rotation during solar minimum.
They observed zones displaying particularly low rotation rates (or high rotation periods) that are likely to be located near the coronal hole/streamer boundaries (\emph{cf.} their Figs. 5 and 6).
Similar observations performed during solar maximum \citep{mancuso_differential_2011} indicate a flatter (more solid-body like) rotation profile, albeit with sub-structures that are harder to link clearly to the coronal topology.s

In any case, a detailed picture of how the solar rotation establishes throughout the corona, including coronal holes and streamers, and whether it can produce sustained wind shears is still lacking. 
Closed magnetic loops will tend toward uniform rotation rates all along them \citep[thus opposing rotation-induced shearing;][]{grappin_mhd_2008}, while open field lines will develop a wind flow that will see its azimuthal speed decreasing with distance from the Sun (in order to conserve angular momentum, as long as the magnetic tension exerted by the background field becomes weak enough).
Streamers are systems of closed magnetic loops, and as such should acquire a shape and rotation pattern that depend on the specific range of solar latitudes that they are magnetically rooted at.
Large streamers can encompass a large range of magnetic loop sizes and footpoint rotational speeds, and may develop an internal differential rotation structure.
Open field lines will follow either a vertical path or one with strong inflexions around streamers depending on where they are rooted at the surface of the Sun, thus having an impact on the transport of angular momentum from the surface up to the high corona.
Thus, rotation shearing layers can develop at specific places of the solar corona, such as at the interfaces between coronal holes and streamers.
Such shear layers can be off importance to the formation of magnetic field reversals, if ever they can develop MHD instabilities that allow for the transport of mass, vorticy and helicity across different topological regions \citep[via shearing, resistive or Kelvin-Helmhöltz instabilities; cf. e.g.][]{dahlburg_modelling_2003, ruffolo_shear-driven_2020}. Additionally, the shear patterns formed can be of importance to the transport and amplification of such magnetic structures (or at least a fraction of them).

The first four PSP perihelia occurred during solar minimum, between about November 2018 and February 2020.
During these close passes, the spacecraft remained within 5 degrees of the solar equator during that time, and also close to the heliospheric current sheet (HCS).
Solar minimum conditions translated into a corona displaying a large equatorial streamer and two polar coronal holes, essentially in an equator-symmetric configuration, except for the occurrence of a small low-latitude CH (visible during the first and second orbits) and a few equatorward polar CH extensions and small equatorial CHs.

This manuscript focuses on investigating the response of the coronal magnetic field and solar wind flows to photospheric rotation, and on pointing out consequences to the interpretation of recent measurements made by PSP.
We used an MHD numerical model of the solar wind and corona, and estimations of the sun-to-spacecraft connectivity using the IRAP's connectivity tool to determine the coronal context of the solar wind flows probed by PSP during its first few solar encounters \rev{(focusing on the first, second and fourth encounters, due to the lack of continued good quality solar wind data for the third encounter)}.
\rev{We suggest that the global dynamics of the rotating solar corona can impact the conditions for the formation of magnetic disturbances such as SBs (among others).}


\section{Numerical model of the rotating corona and wind}
\label{sec:dip}

\begin{figure*}[!t]
  \centering
  \includegraphics[clip,trim=0 50 0 50, width=0.32\linewidth]{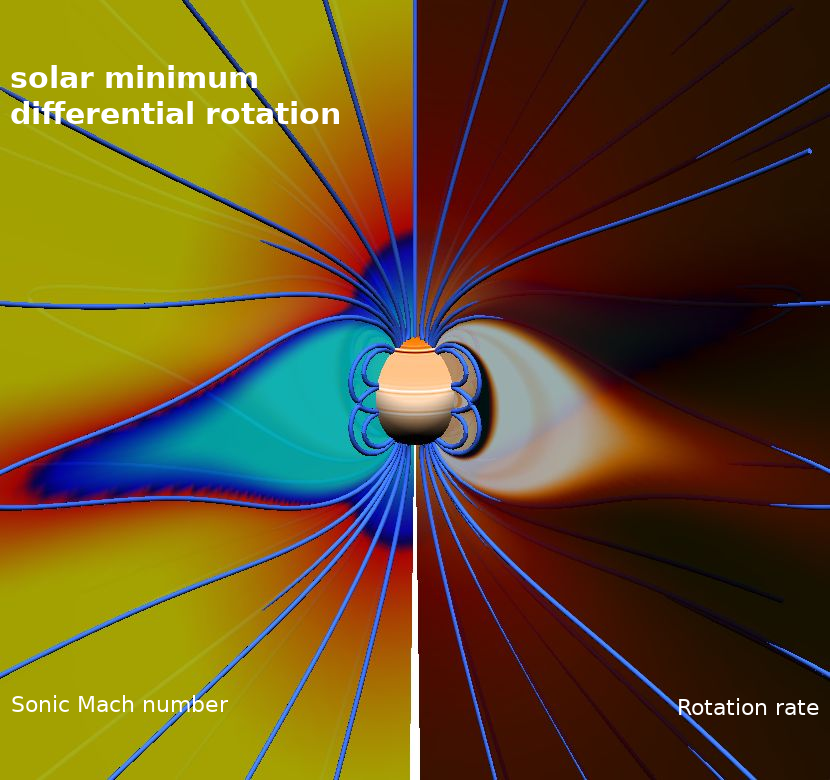}
  \includegraphics[clip,trim=0 50 0 50, width=0.32\linewidth]{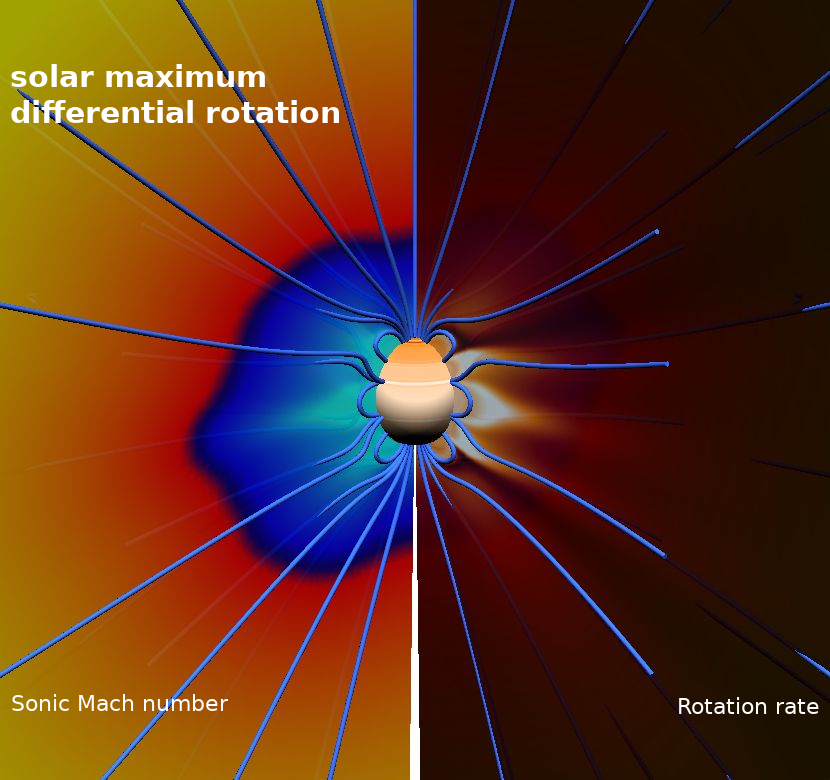}
  \includegraphics[clip,trim=0 50 0 50, width=0.32\linewidth]{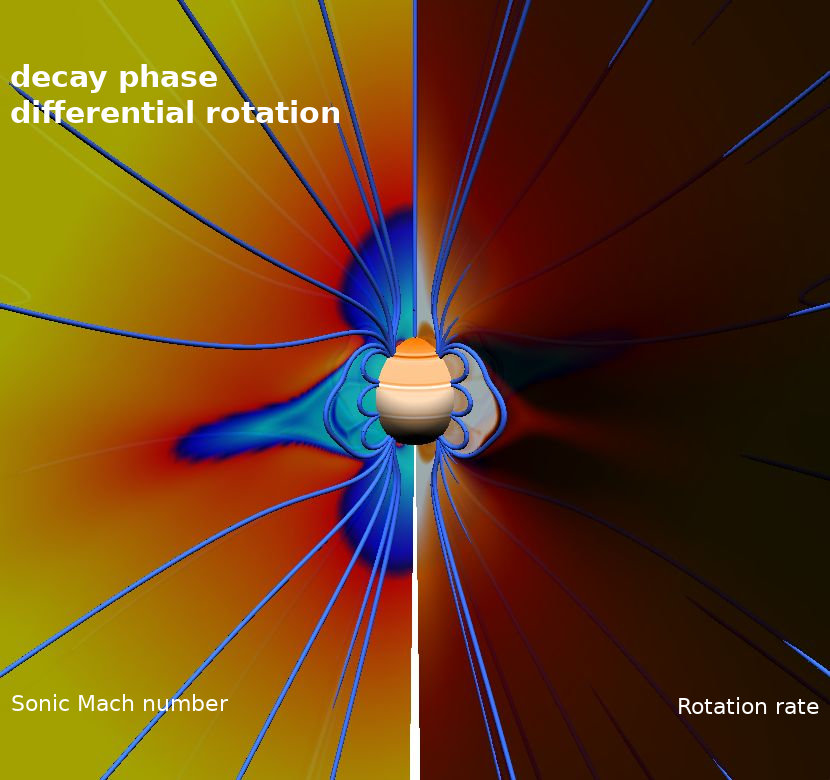}
  
  \includegraphics[clip,trim=0 50 0 50, width=0.32\linewidth]{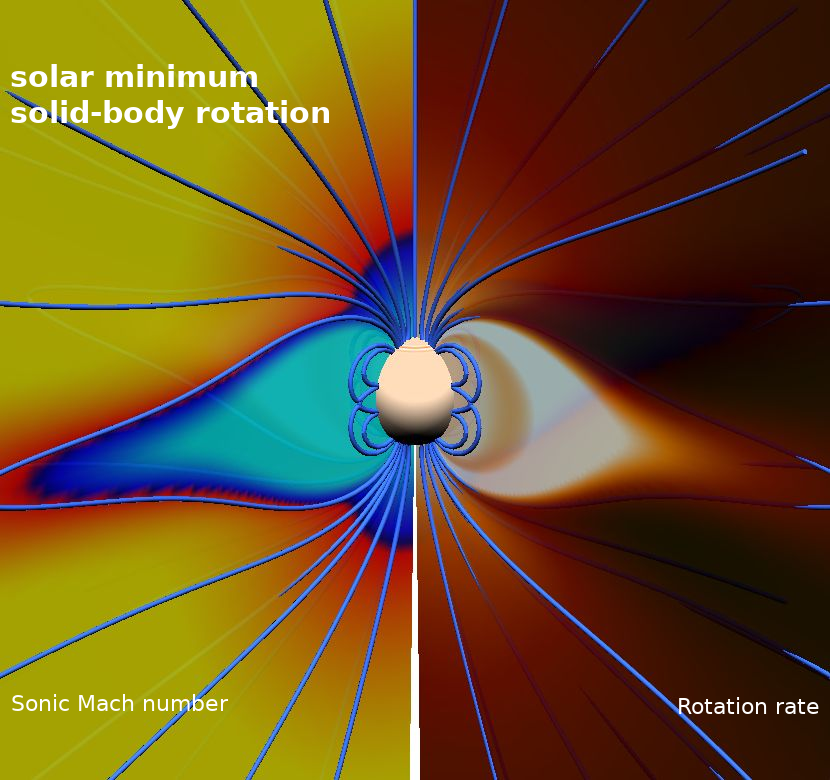}
  \includegraphics[clip,trim=0 50 0 50, width=0.32\linewidth]{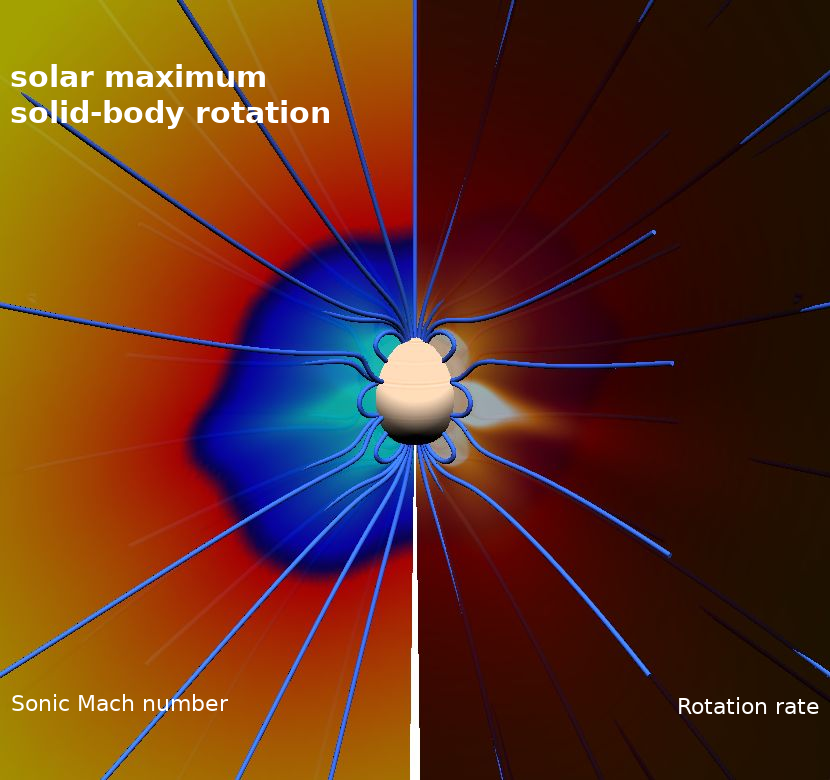}
  \includegraphics[clip,trim=0 50 0 50, width=0.32\linewidth]{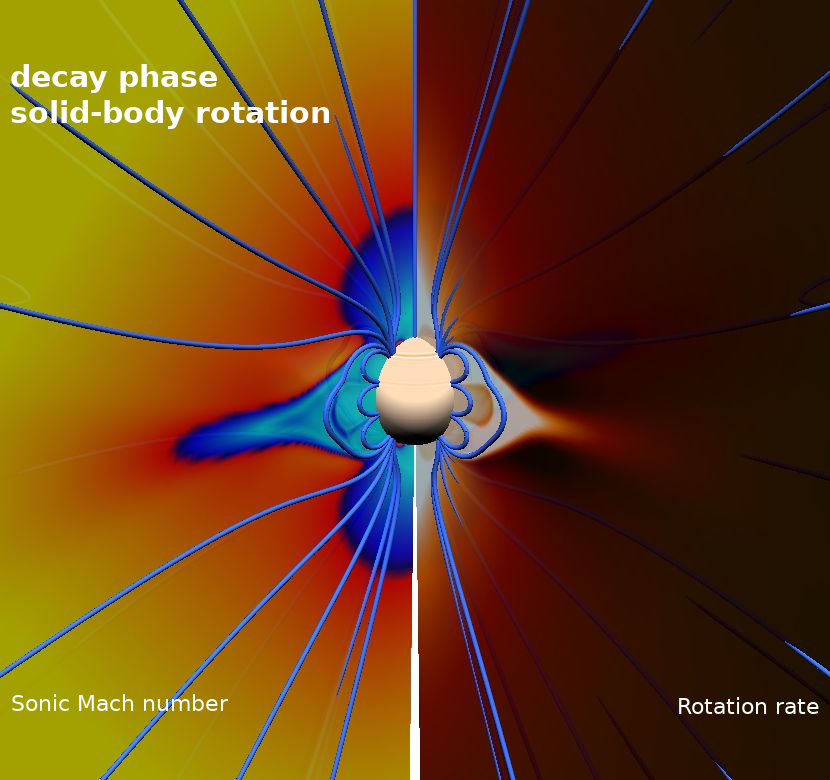}

  \caption{
    Coronal magnetic field, wind velocity and rotation rate at three illustrative moments of the solar cycle (from left to right: solar minimum, maximum and decay phase, corresponding to instants $t=0$, $3.8$, and $4.\un{yr}$ in Fig. 3 of \citep{pinto_coupling_2011}).
    The left halves of the panels show meridional slices of the wind speed (sonic Mach number, from $0$ in solid blue to $2$ in solid yellow, and with the yellow/blue boundary representing the sonic surface), and the right halves show slices of the rotation rate of the solar corona (from $0$ in black to $14\un{^\circ/day}$ in light orange).
    Top and bottom rows correspond to cases with a standard differential rotation and to solid-body rotation at the lower boundary (the lower boundary of the corona is coloured with the same $\Omega$ colour scale as the slices on the right side of the images).
    Blue lines represent magnetic field lines.
    The frame of reference is inertial (not comoving with the Sun), hence $\Omega$ is positive everywhere. 
  }
  \label{fig:mach_omega_panels}
\end{figure*}

\begin{figure*}[]
  \centering

  \includegraphics[clip,trim= 0 23 0 0, height=0.19\linewidth]{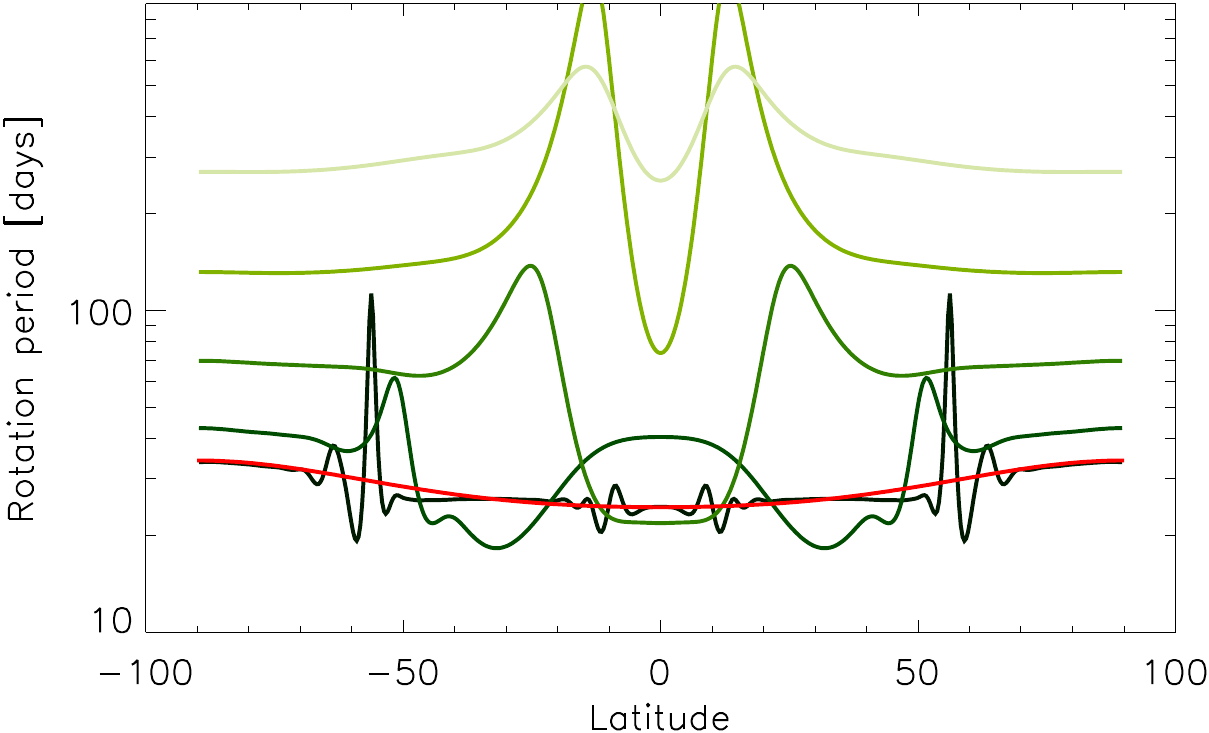}
  \includegraphics[clip,trim=41 23 0 0, height=0.19\linewidth]{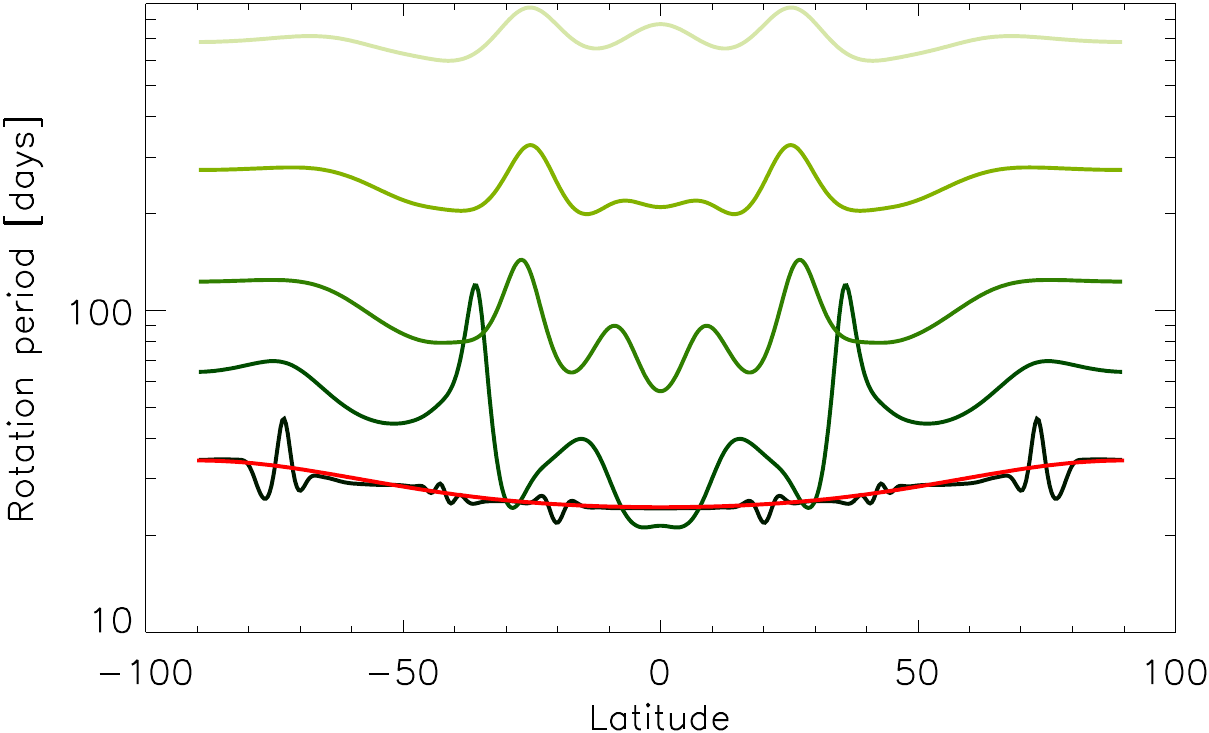}
  \includegraphics[clip,trim=41 23 0 0, height=0.19\linewidth]{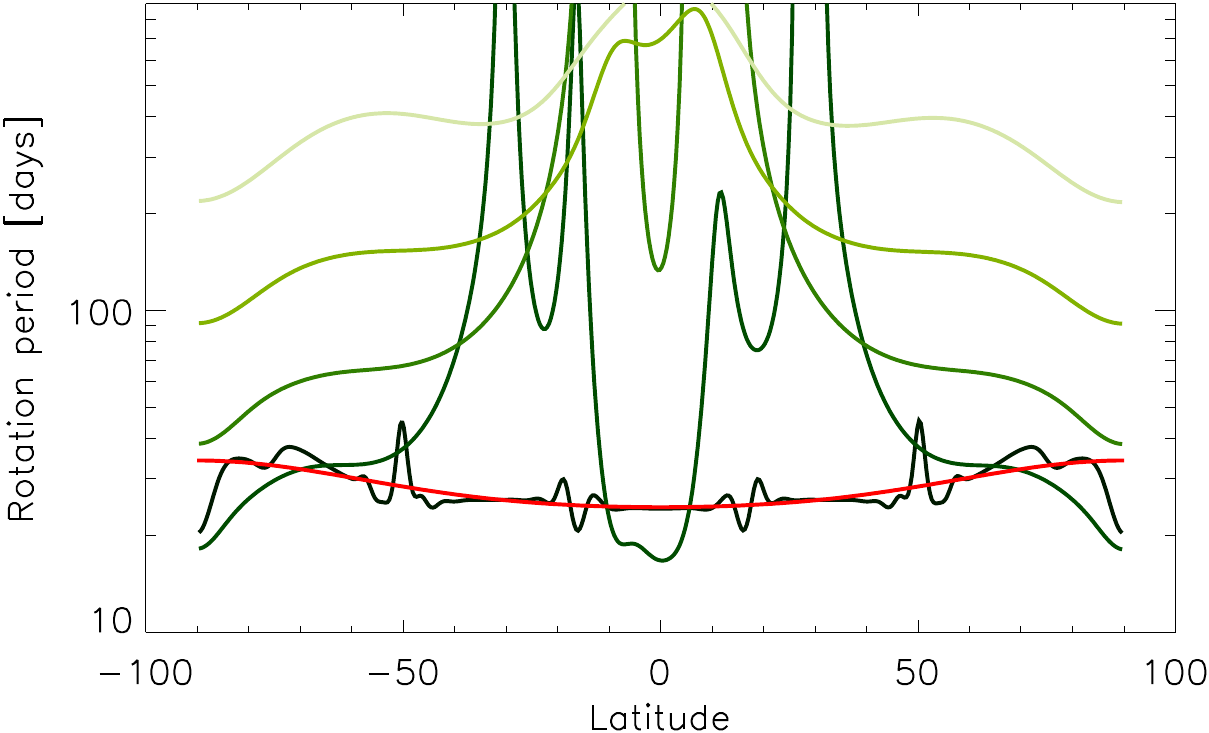}

  \includegraphics[clip,trim= 0 0 0 0, height=.2135\linewidth]{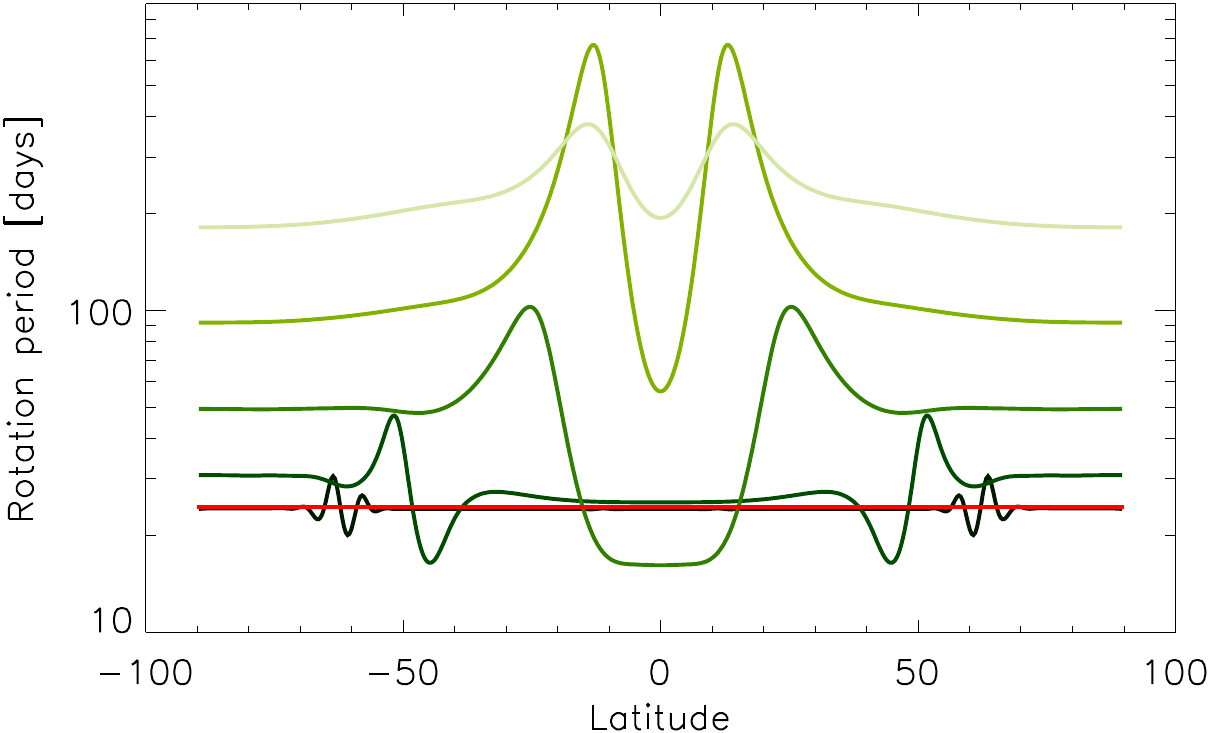}
  \includegraphics[clip,trim=41 0 0 0, height=.2135\linewidth]{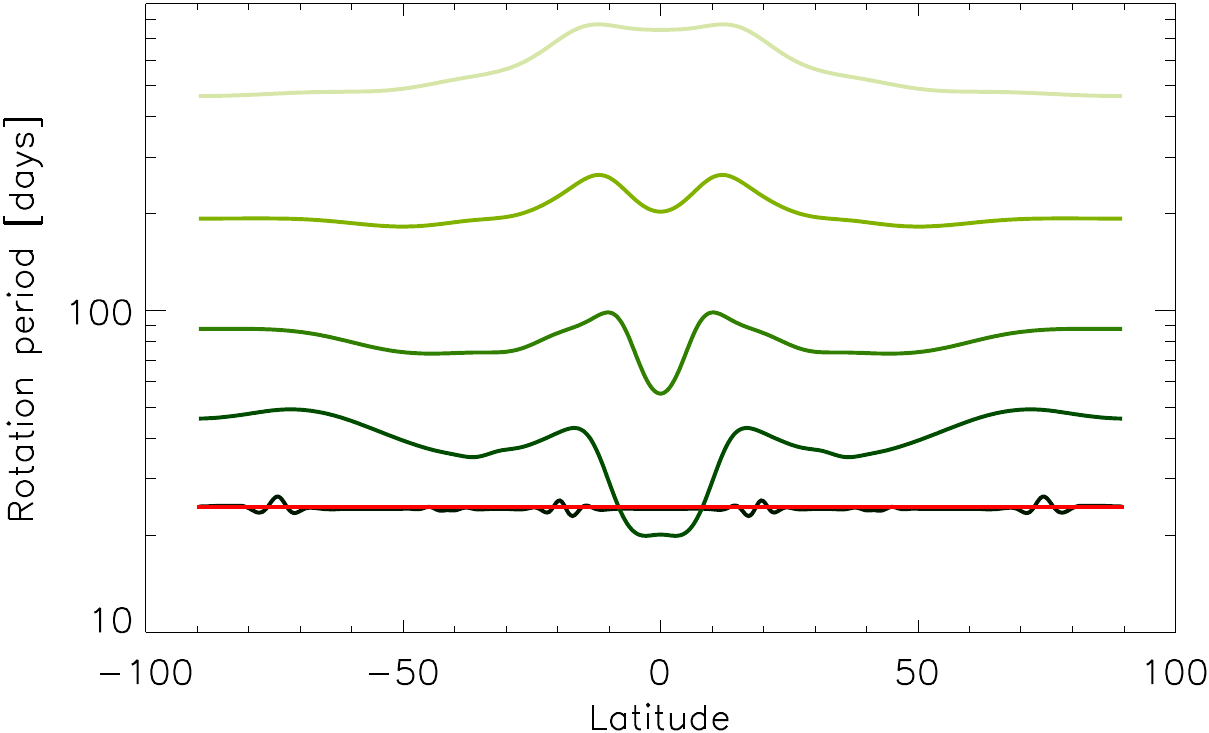}
  \includegraphics[clip,trim=41 0 0 0, height=.2135\linewidth]{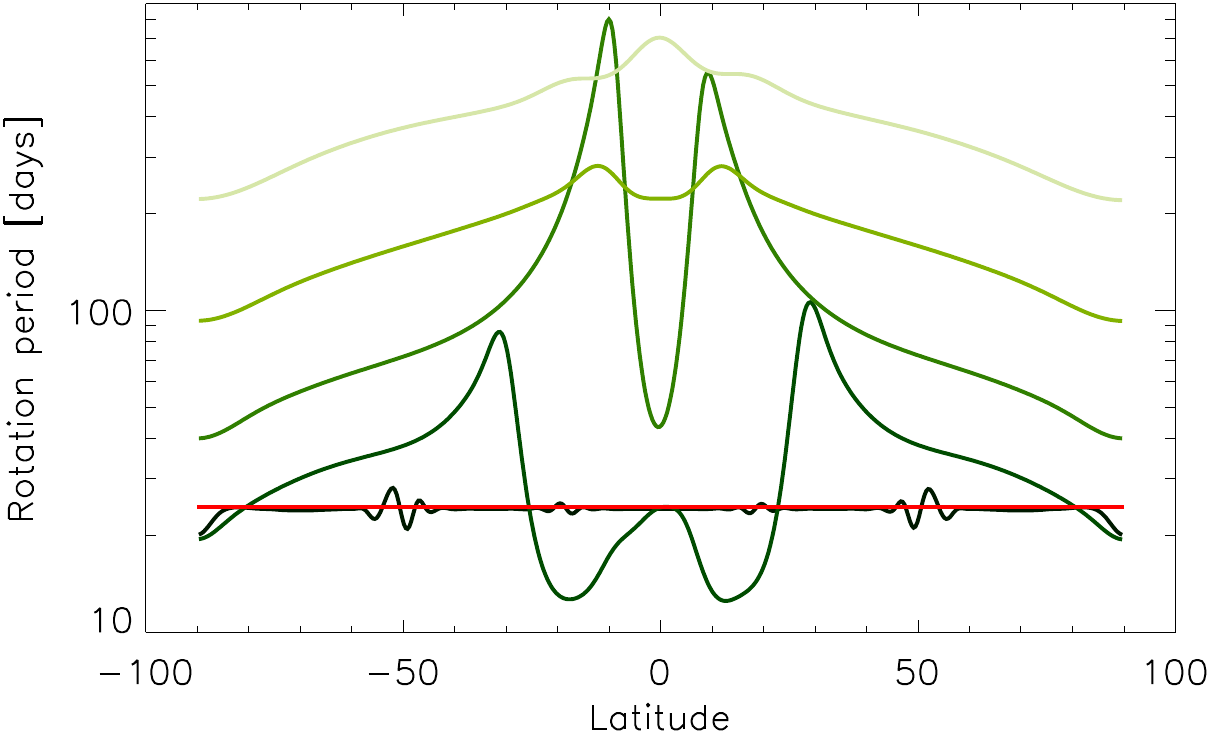}

  \caption{
    Rotation period as a function of latitude at different solar radii ($r = 1.03,\ 2.0,\ 4.0,\ 8.0$ and $16 \rsun$, from darker to lighter green lines) for the runs represented in Fig. \ref{fig:mach_omega_panels} (from left to right: solar minimum, maximum and decay phase; differential surface rotation on the top row, solid-body rotation on the bottom row).
    The red curves indicate the rotation period imposed at the surface (eq. \ref{eq:rotation_rate}, with $\Omega_b$ and $\Omega_c$ equal to $0$ in the solid rotation case).
    Rotation periods peak just outside CH/streamer boundaries, as in the UVCS observations by \citet{giordano_coronal_2008}.
    Global (resonant) oscillations of closed loops within the streamers are visible within these main peaks, especially in the case with differential rotation.
    Maximum rotational shearing occurs at mid-altitudes (below maximum streamer height).
    Wind shear is transmitted upwards, well above the streamer heights, at the vicinity of HCSs.
    At higher heights, the corona tends to progressively approach solid-body rotation with height within coronal holes (note that the case in the third column has a polar pseudo-streamer).
  }
  \label{fig:rot_periods_plots}
\end{figure*}

\begin{figure*}[!h]
  \centering

  \includegraphics[width=0.45\linewidth]{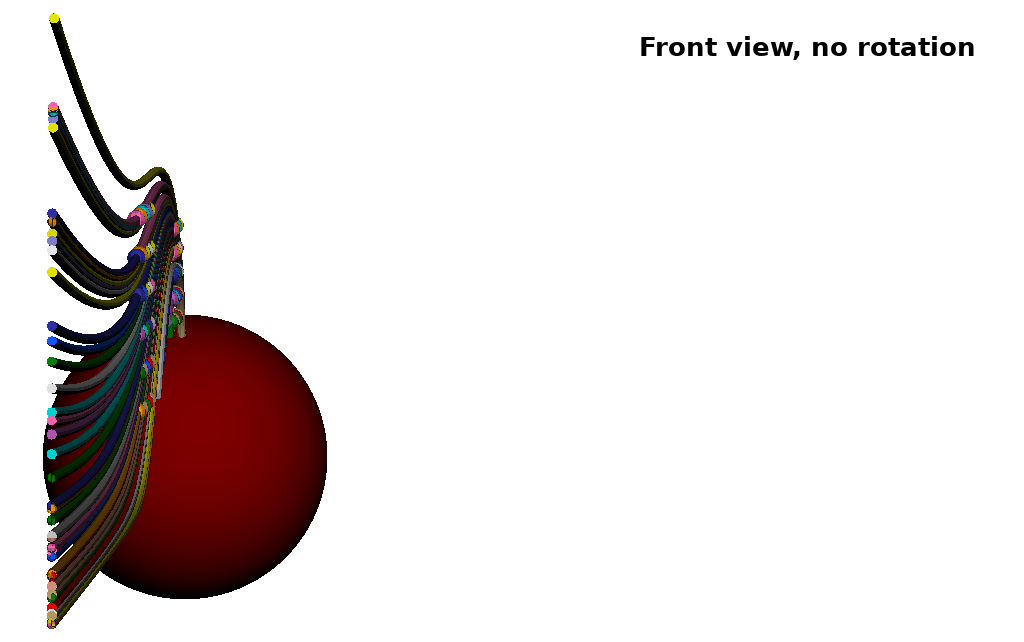}
  \includegraphics[width=0.45\linewidth]{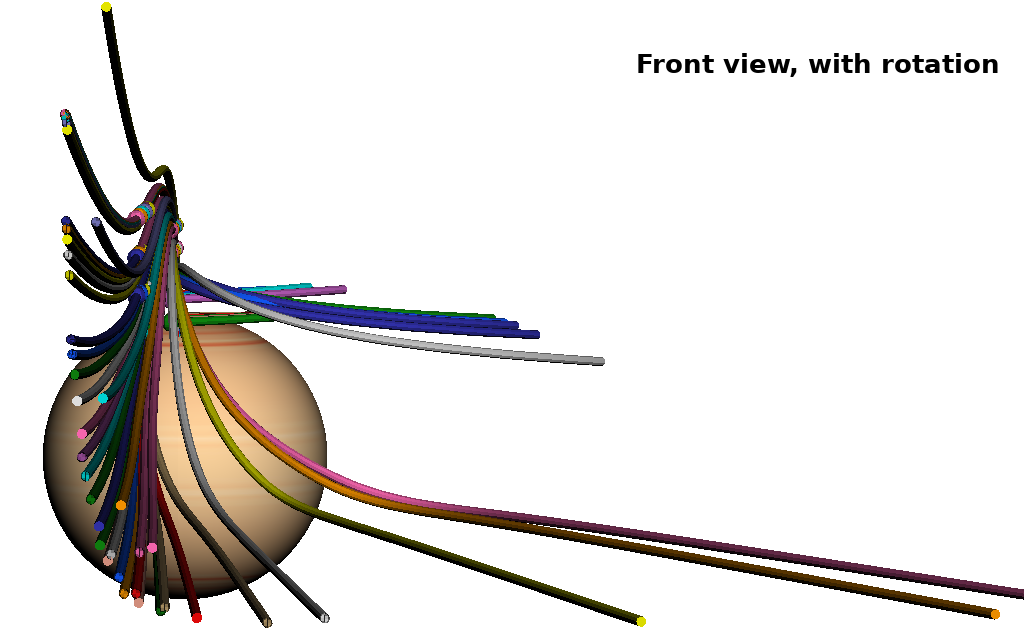}

  \includegraphics[width=0.45\linewidth]{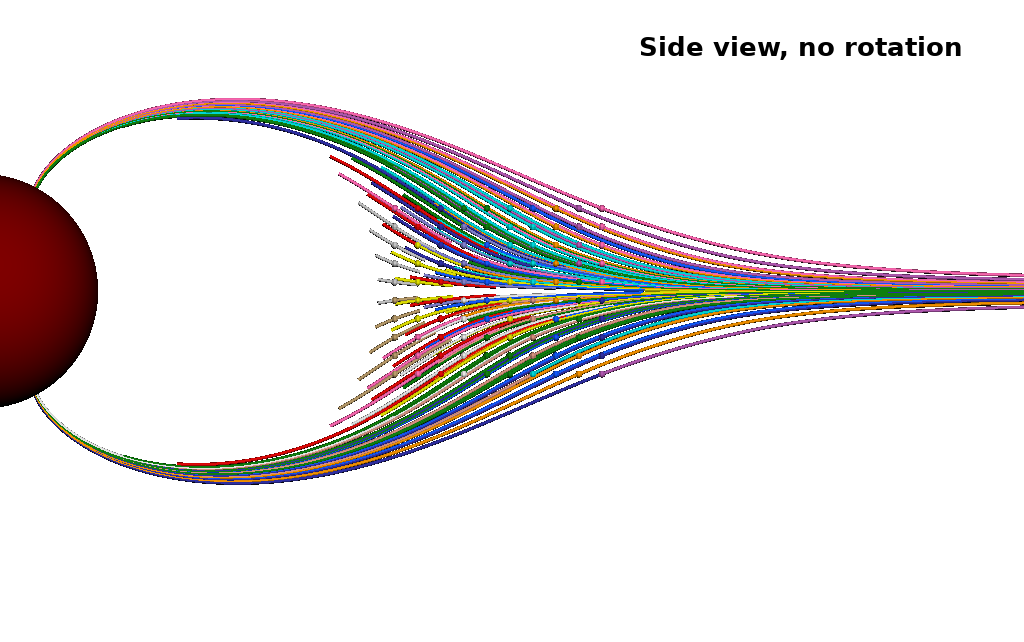}
  \includegraphics[width=0.45\linewidth]{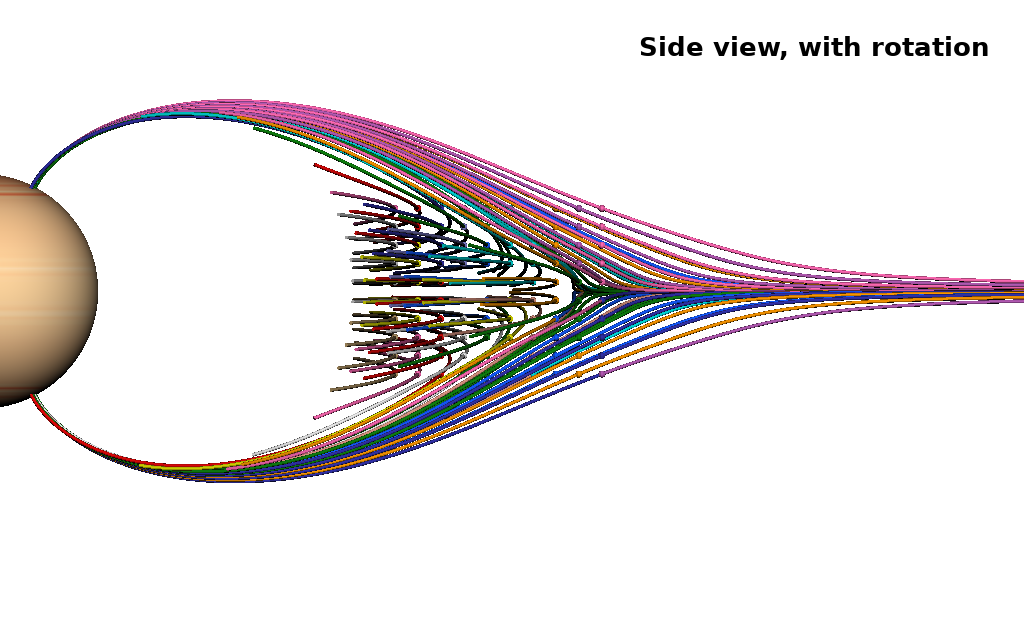}

  \includegraphics[width=0.45\linewidth]{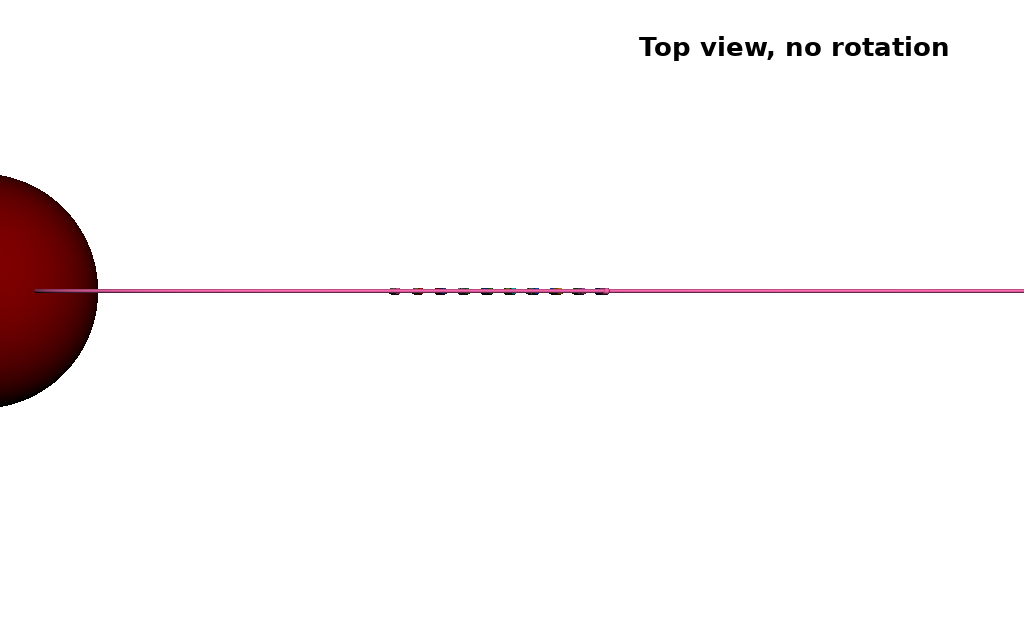}
  \includegraphics[width=0.45\linewidth]{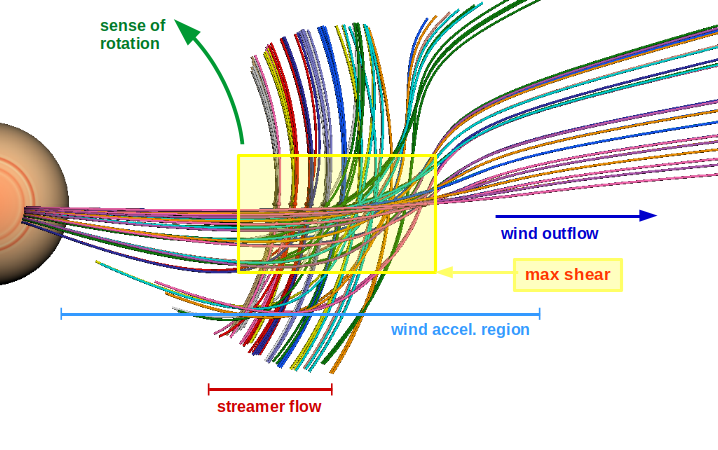}

  \caption{
    Flowlines for the non-rotating (left) and rotating (right) cases, rendered from different perspectives (from top to bottom: front, side, and pole-on views).
    For visual clarity, only the streamlines crossing the northern hemisphere are shown in the front view.
    The length of each streamline indicates the total displacement of a given plasma element during the same period of time.
    Streamlines that correspond to plasma lying well within the streamer make an arc of a circle around the Sun (CCW on the pole-on view), while those connected to solar wind flows extend outward.
  }
  \label{fig:flowlines_main}
\end{figure*}

\subsection{Numerical code and setup}
\label{sec:dip_setup}

We used the numerical code DIP to model a 2.5 D axi-symmetric solar corona, setup in a similar way as in \citet{pinto_flux-tube_2016} and \citet{pinto_coupling_2011}, although with a higher spatial resolution ($768\times 768$ grid, non-uniform in radius and uniform in latitude) and including rotation.
The code solves a system of MHD equations that describes a one-fluid, isothermal, fully ionised and compressible plasma:
\begin{eqnarray}
  \label{eq:mhd_dip}
  \partial_t \rho & + & \diver{\rho\mathbf{u}} =0\ , \label{eq:continuity} \\
  P  &=& \frac{2}{m_H}\rho k_BT \ , \label{eq:pressure} \\
  \partial_t \mathbf{u} & + & \left(\mathbf{u}\cdot\nabla\right)
  \mathbf{u} = -\frac{\nabla P}{\rho} +
  \frac{\mathbf{J}\times\mathbf{B}}{\mu_0\rho} -
  \mathbf{g} + \nu\nabla^2\mathbf{u} \ , \label{eq:momentum} \\ 
  \partial_t\mathbf{B} &=& \rot{\left(\mathbf{u}\times\mathbf{B}\right)} +
  \eta\nabla^2\mathbf{B}\ . \label{eq:induction}
\end{eqnarray}
The model assumes the corona and the solar wind to be isothermal with a uniform coronal temperature $T_0 = 1.3 \un{MK}$ and a specific heat ratio $\gamma = 1$.
The magnetic field $\mathbf{B}$ separates into a time-independent external component $\mathbf{B^0}$ (a potential field resulting from the internal dynamics of the Sun) and into an induced field $\mathbf{b}$.
We adopted several configurations for $\mathbf{B^0}$ in order to simulate different moments of the solar activity cycle.
The equations are integrated using a high-order compact finite difference scheme \citep{lele_compact_1992} with third-order Runge-Kutta time-stepping \citep[cf.][]{grappin_alfven_2000}.
The diffusive terms are adapted to the local grid scale ($\Delta l$), that is non-uniform in the radial direction ($\Delta l$ is minimal close to the lower boundary).
The kinematic viscosity is defined as $\nu=\nu_0\left(\Delta l/\Delta l_0 \right)^2$,
typically with
$\nu_0=2\times 10^{14}\un{cm^2\cdot s^{-1}}$
and
$0.01\lesssim\left(\Delta l/\Delta l_0 \right)^2\lesssim10$.
The magnetic diffusivity $\eta$ is scaled similarly.

The boundary conditions are formulated in terms of the MHD characteristics by imposing the amplitudes of the MHD characteristics propagating into the numerical domain (the outgoing ones being already completely determined by the dynamics of the system).
The upper boundary is placed at $r=15\rsun$ and is fully transparent (open to flows and transparent to waves).
The lower boundary  is placed at $r=1.01\rsun$ and is semi-reflective with respect to the Alfvén mode (but transparent with respect to all others), and is also open to flows.
We treat the chromosphere and the transition region layers as an interface (or rather a discontinuity), and define the chromospheric reflectivity $a$ in terms of the ratio $\epsilon$ of Alfvén wave speeds above and below:
\begin{equation}
  \label{eq:a}
  a = \frac{\epsilon - 1}{\epsilon + 1}\ ,\ \mathrm{with}\ 
  \epsilon = \frac{C_A^{photosph}}{C_A^{corona}}\ ,
\end{equation}
$C_A$ representing the Alfvén speed $B/\left(\mu_0 \rho\right)^{1/2}$.
This approximation is valid for perturbations whose characteristic wavelength is much larger than the thickness of the chromosphere, which is the case for the quasi null-frequency (non-oscillating) rotational forcing that we apply here at the lower boundary.

In order to establish coronal rotation, we first let a non-rotating solar wind solution fully develop in the whole numerical domain, and then apply a torque at the lower boundary which accelerates it progressively to the following rotation rate profile
\begin{equation}
  \label{eq:rotation_rate}
  \Omega\left(\theta\right) = \Omega_a + \Omega_b \sin^2\theta + \Omega_c\sin^4\theta,\ 
\end{equation}
with $\theta$ being the latitude, $\Omega_a = 14.713\un{^\circ/day}$, $\Omega_b = -2.396\un{^\circ/day}$ and $\Omega_c = -1.787\un{^\circ/day}$, following \citet{snodgrass_rotation_1990}.
We also tested solid-body solar rotation profiles, that were achieved by setting $\Omega_b$ and $\Omega_c$ to $0$ in eq. \eqref{eq:rotation_rate}.
The duration of the initial acceleration period was defined to be $\sim1/4$ of the average (final) rotation period at the surface, and we let the system relax for at least $10$ Alfvén crossing times of the whole domain (lower to upper boundary).
The initial transient propagates upwards (after crossing the idealised chromospheric interface) predominantly as an Alfvénic wavefront (with little power on the non-alfvén charateristics), accelerating the open field plasma in the azimuthal direction and exciting a few global oscillations in the closed field regions.
After a few Alfvén transit times, and for a small enough $\epsilon$, the corona and wind settle down into a quasi-steady state.
For $\epsilon\sim 1$, the highly transparent surface spins down very quickly, as this configuration corresponds to a lower boundary that cannot oppose the braking torque that results from net outward angular momentum flux carried away by the solar wind.
For $\epsilon\sim 0$, the lower coronal boundary maintains the imposed rotation, and the magnetic field is line-tied to it.
Smaller closed loops keep oscillating resonantly for a long time and the larger ones are sheared indefinitely (as their foot-points suffer a larger range of azimuthal speeds due to differential surface rotation).
Intermediate (more solar-like) values of $\epsilon$ allow for the surface rotation to be maintained in the open-flux regions while the minimal required amount of footpoint leakage is allowed for the closed-flux regions to stabilise \citep[\emph{cf.}][]{grappin_mhd_2008}.
The streamers gradually evolve towards a nearly solid-body rotation profile, with a rotation rate determined by that of its magnetic foot-points.
Large streamers encompass a wide latitudinal range, and therefore a wide range of footpoint rotation rates.
As a result, such streamers develop a more distinguishable differential rotation pattern than the smaller ones.
The open field regions (coronal holes) develop permanent azimuthal velocity and magnetic field components, with a reasonably complex spatial distribution in the low corona converging into what can be thought of as the beginning of the Parker spiral on the outer part of the domain.
The finite magnetic resistivity affects the width of these regions, but has a negligible effect on the overall rotation rates.
We note that in order to achieve long-lasting (\emph{i.e}, stable) coronal rotation profiles we had to resort to values of $\epsilon$ of about $10^{-3}$, rather than to a more realistic $\epsilon=10^{-2}$.
The reason for this is that the convective dynamics of the surface and sub-surface layers of the Sun are absent from our model, and therefore so are the resulting torques that would counter-balance the weak (but finite) braking torque exerted on the lower boundary by the rotating solar wind.

s

\subsection{Rotating corona and solar wind: overview}
We ran a series of numerical MHD simulations of the solar corona and wind on which axisymmetric streamers and coronal holes are set under rotation following the methods described in Sect. \ref{sec:dip}.
Figure \ref{fig:mach_omega_panels} shows three-dimensional renderings of different simulation runs, corresponding to different moments of the solar cycle (from left to right: activity minimum, maximum and decay phase) and to different surface rotation profiles (from top to bottom: differential and solid-body).
The blue lines are magnetic field lines.
The left halves of images display meridional cuts of the solar wind speed (in units of sonic Mach number, from $0$ in solid blue to $2$ in solid yellow), while the right side halves show the rotation rate $\Omega$ (from $0$ in black to $14\un{^\circ/day}$ in light orange, defined \rev{with} respect to the inertial reference frame).
The solar surface is coloured with the same colour scheme as the right side of the images, and hence show the rotation pattern at the lower coronal boundary.
It is immediately clear from the figure that the solar corona assumes a rotation state that is highly structured and that reflects the large-scale topology of the magnetic field.
As described above, the streamers tend to set themselves into solid body rotation, with the largest ones developing a more complex internal rotation structure (with some of its larger inner loops undergoing global resonant transverse oscillations for a long period of time).
Open field lines that pass well within coronal holes (faraway from CH/streamer boundaries) progressively develop a backward bend (opposed to the direction of rotation).
The open field-lines that pass close to the coronal hole boundaries suffer stronger expansions and deviations from the radial direction, drive the slowest wind flows, and acquire the lowest rotation speeds at mid coronal heights (not very far from the streamer tips).
The rotation rates remain high at and right above the streamers tips (close to the HCS/HPS), nevertheless, as these tend to assume a rotation profile more closely linked with that of the surface (and of the corresponding streamers).
As a result, the strongest spatial contrasts in wind speed and rotation rate $\Omega$ are usually found across these regions of the corona.

\begin{figure*}[!t]
  \centering

  Wind speed gradient $\left| \nabla u \right|$ \\
  \includegraphics[clip,trim=150 120 240 218, height=0.255\linewidth]{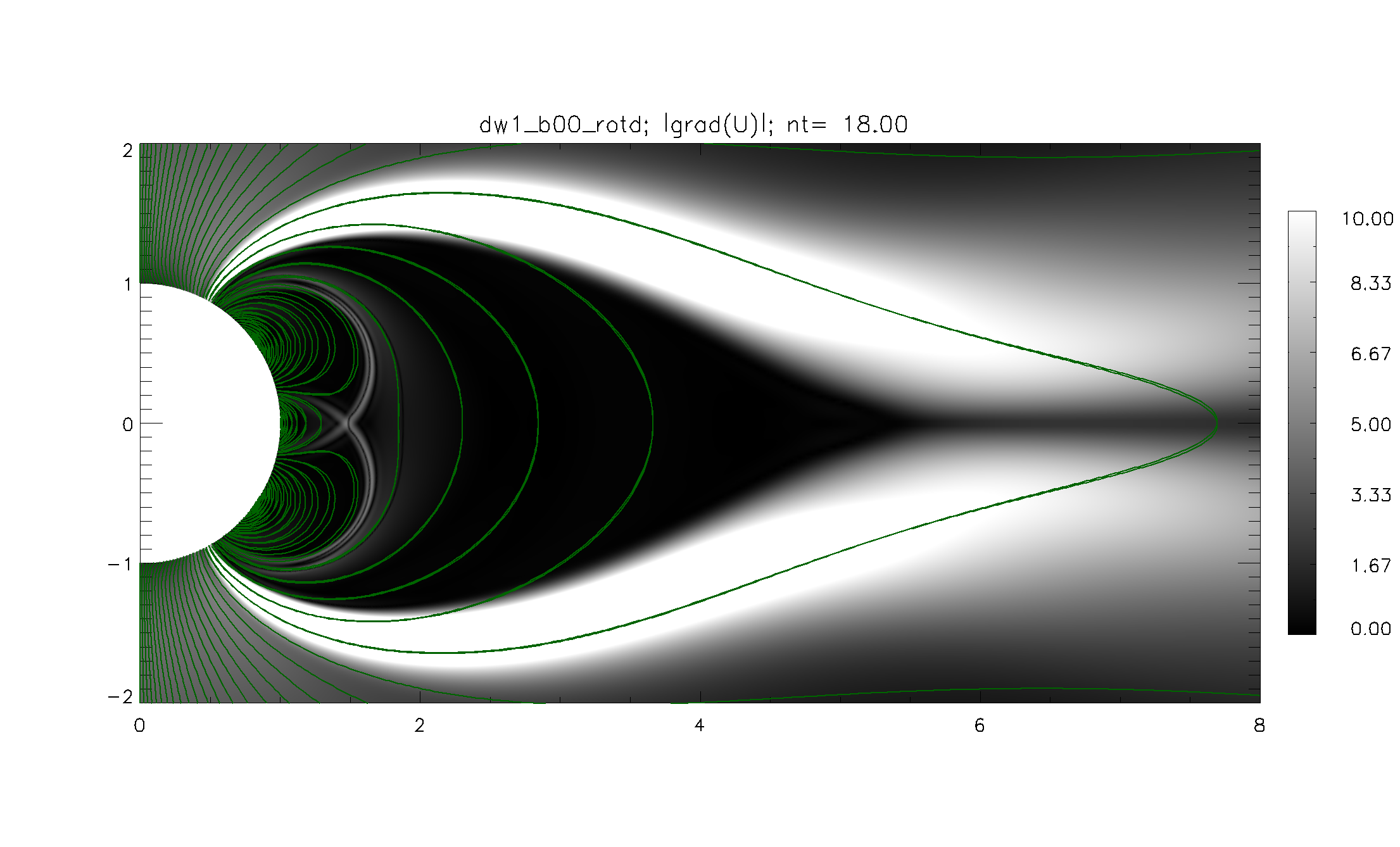}
  \includegraphics[clip,trim=150 120 5   218, height=0.255\linewidth]{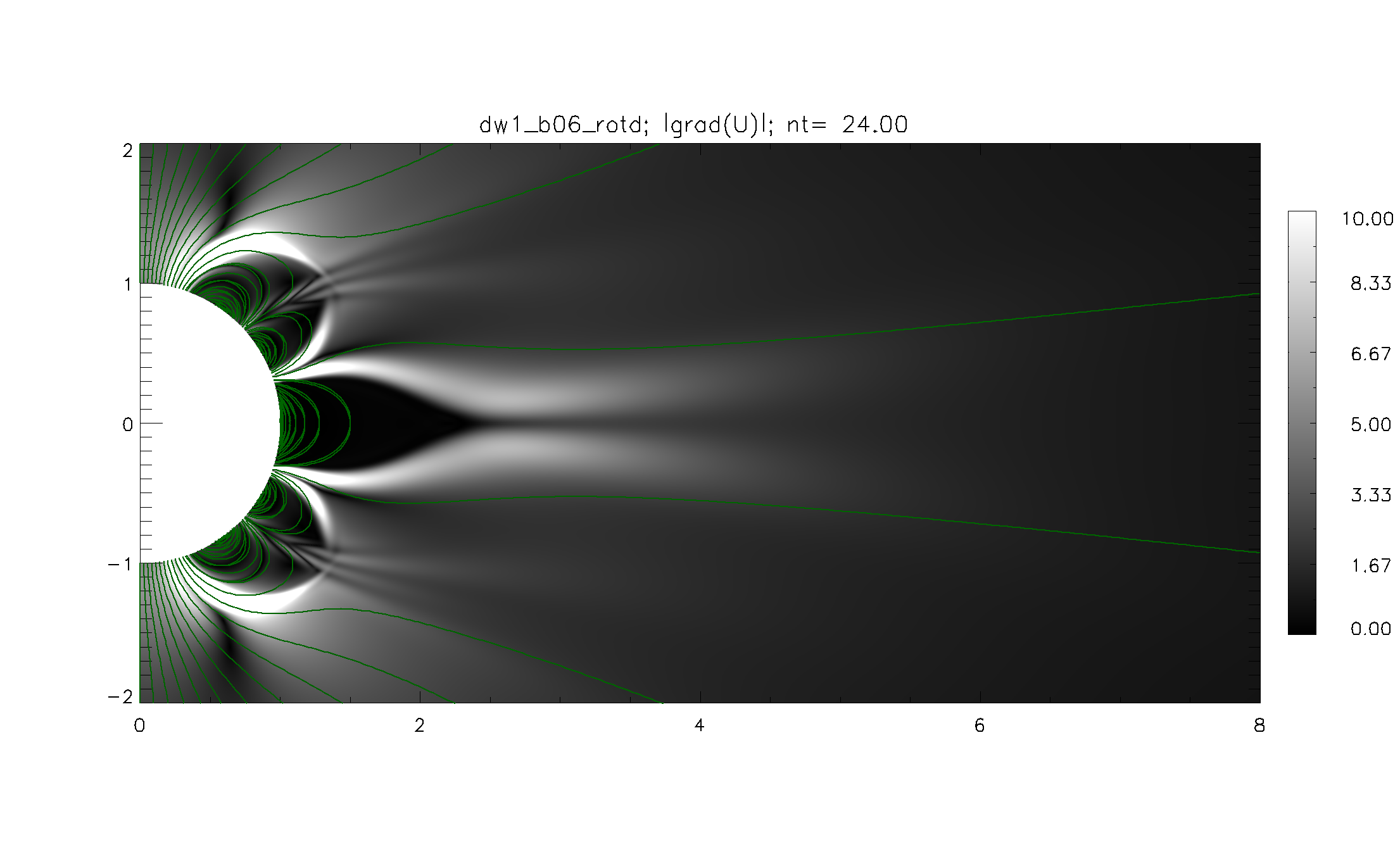}

  Rotation rate gradient $\left| \nabla \Omega \right|$ \\
  \includegraphics[clip,trim=150 120 240 218, height=0.255\linewidth]{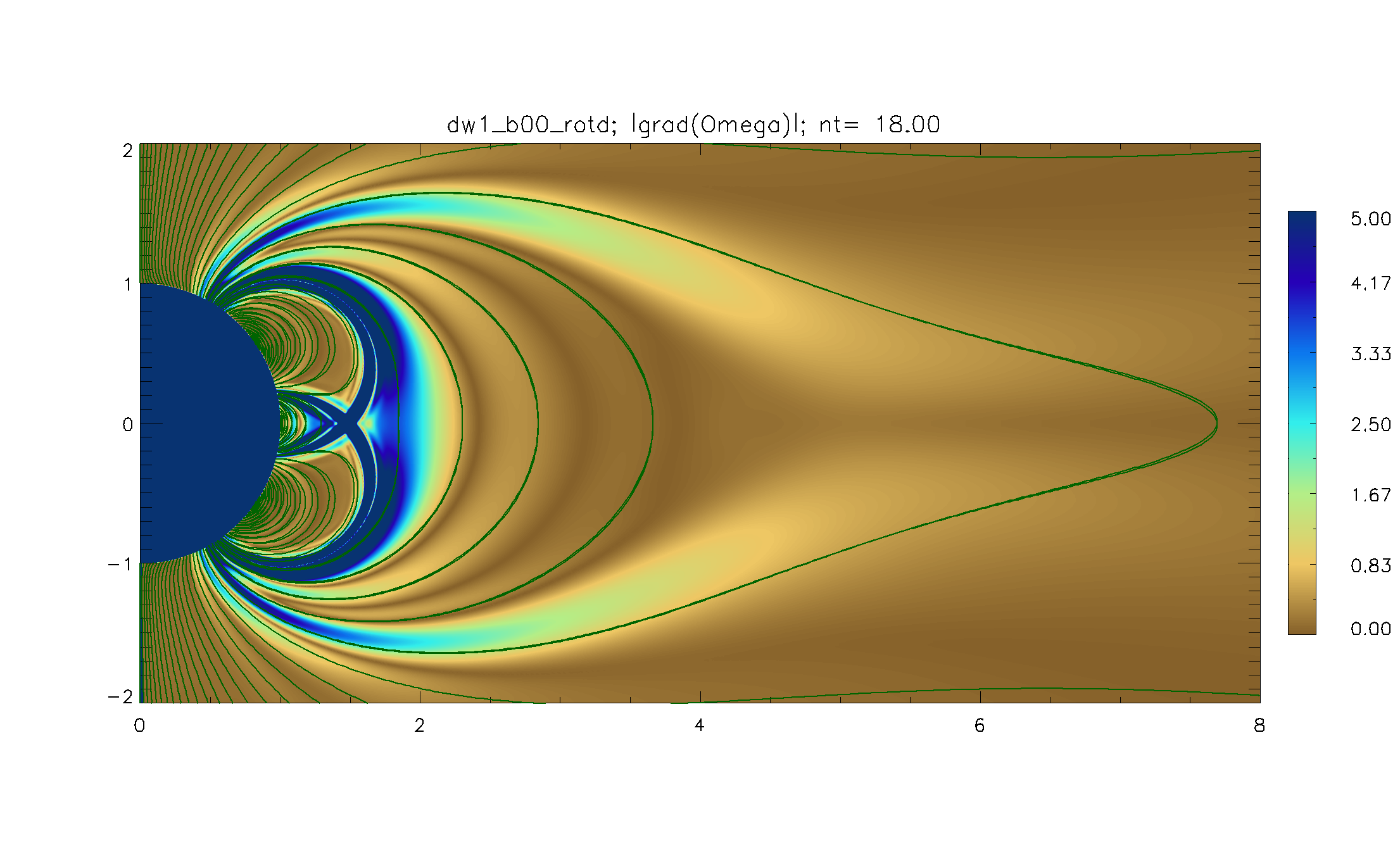}
  \includegraphics[clip,trim=150 120 5   218, height=0.255\linewidth]{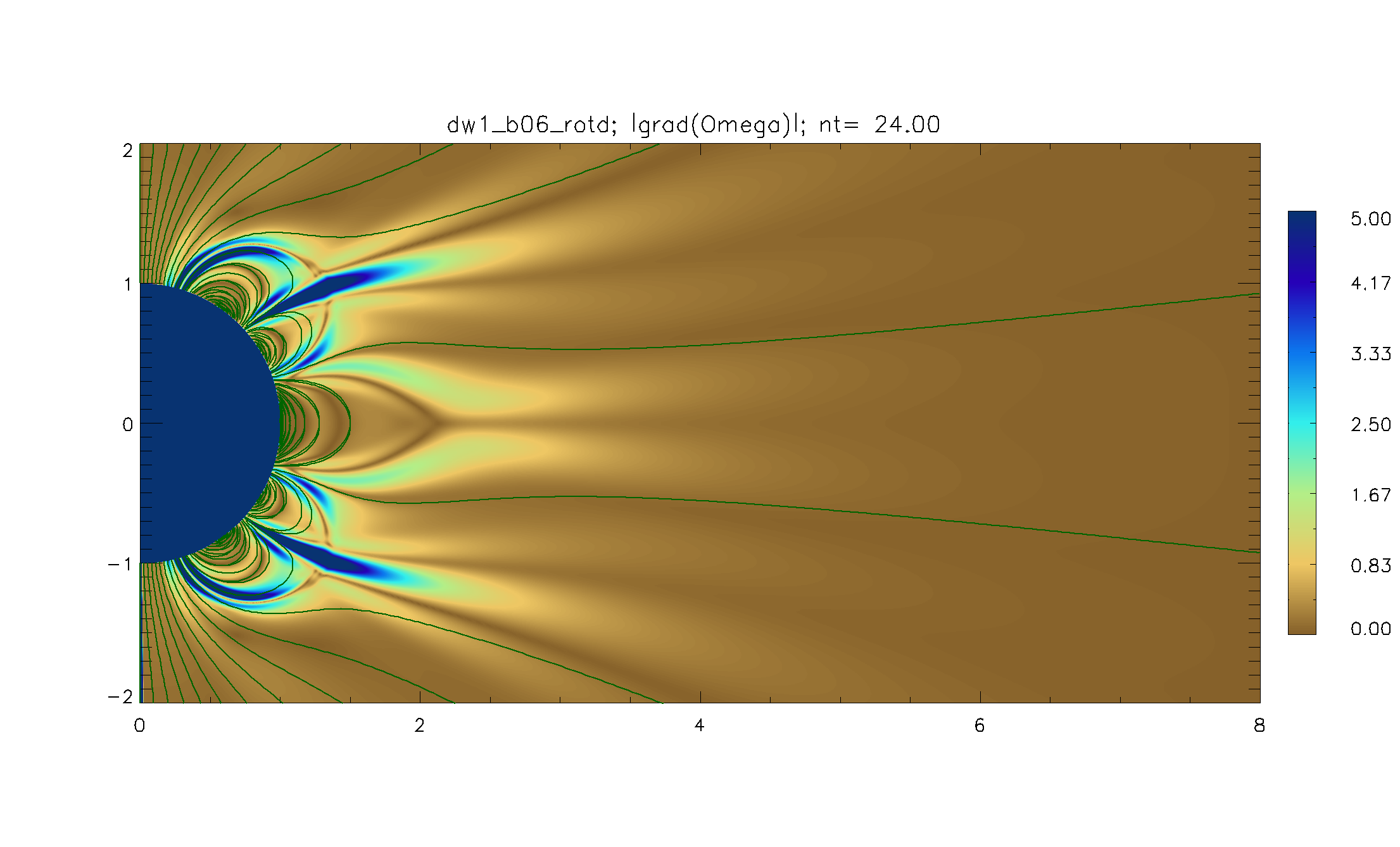}

  \caption{
    Zoomed-in view of the absolute magnitudes of the gradients of flow speed (top) and of the rotation rate $\Omega$ (bottom) close to the equatorial streamer on the solar minimum (left) and maximum (right) configurations.
    For simplicity, only differentially-rotating cases are shown and both quantities are displayed in normalised units (in units of $6.27\e{-4}\un{s^{-1}}$ and $9.0\e{-13}\un{rad\ s^{-1} m^{-1}}$, respectively).
    The axis indicates distances in solar radii. 
    The green lines are magnetic field-lines.
    The coronal hole/streamer boundaries systematically develop flow shears due to gradients in the solar wind flow along the magnetic field and to gradients in the rotation pattern (flow across the field) altogether.
    These shearing regions reach well into the coronal holes, and into the wind flow well past the streamer top height.
  }
  \label{fig:flow_grad}
\end{figure*}
\subsection{Varying surface rotation profiles and cycle phase}

Different surface rotation profiles lead to qualitatively similar results,
with differences between the differentially-rotating and the solid-body cases being more pronounced on large coronal loop systems. 
Streamers that are fully rooted at mid or high-latitude (i.e, not equator-symmetric) are subject to footpoint shearing (albeit shallow in amplitude), and acquire an orientation oblique in respect to the meridional plane (which does not happen with solid-body rotation).
The positions and amplitudes of the fast and slow wind streams do not change when switching between the two types of rotation.
The positions of the higher and lower rotation rates regions are also maintained, but their amplitudes and substructure can differ sensibly.

Figure \ref{fig:rot_periods_plots} shows the rotation period (in days) as a function of latitude at different radii ($r = 1.03,\ 2.0,\ 4.0,\ 8.0$ and $16 \rsun$, from darker to lighter green lines) for the same runs as those represented in Fig. \ref{fig:mach_omega_panels} (from left to right: solar minimum, maximum and decay phase; differential surface rotation on the top row, solid-body rotation on the bottom row).
The red curves indicate the rotation period imposed at the surface (eq. \ref{eq:rotation_rate}, with $\Omega_b$ and $\Omega_c$ equal to $0$ in the solid-body rotation case).
The rotation regimes of coronal holes and closed-field regions are clearly distinct, with the former tending to progressively approach solid-body rotation with increasing altitude (note that the case in the third column has pseudo-streamers at the north and south poles), and the latter showing a more complex rotation structure.
The direct imprint of the imposed surface profile is only observed within the first few solar radii, with the overall latitudinal trend of the rotation period (equator to poles) even reversing in the high corona in some cases (e.g., in the first and third columns of the figure).
Remarkably, rotation periods peak strongly just outside CH/streamer boundaries.
These peaks are present along the whole extension of the boundaries (note how they get closer together with increasing altitude, especially for the cases with large streamers), and even beyond.
They correspond spatially to the slowing down of the rotation rate visible as dark patches in the right halves of the panels in Fig. \ref{fig:mach_omega_panels}.
These features are particularly prominent in the solar minimum configuration, with a very large equatorial streamer, and strikingly consistent with the solar coronal rotation periods measured below $2\rsun$ by \citet{giordano_coronal_2008} using SoHO/UVCS data (during activity minimum, May 1996 to May 1997).
Their observations highlight that, as in our simulation, the larger gradients of rotation rate are found at the boundaries between open and closed magnetic field lines.
Our simulations suggest furthermore that this behaviour is universal across the activity cycle, although less easy to distinguish during solar maximum (cf. second column in Fig. \ref{fig:rot_periods_plots}), due to a more intricate mixture of smaller streamers/pseudo-streamers and of thinner coronal holes.
This solar minimum to maximum variation is also consistent with the results of \citet{mancuso_differential_2011}, who performed a similar analysis \rev{with} UVCS data during solar maximum (March 1999 to December 2002).
Global (resonant) oscillations of closed loops within the streamers are visible in the interval between these main peaks in rotation period (that is, within streamers), especially in the cases with differential rotation.
While latitudinal gradients in rotation rate tend to be maximal at mid-altitudes (below maximum streamer height), they remain significant much beyond the height of the largest streamers, at the vicinities of HCSs.
The gradients in rotation period are overall steeper in the cases under differential surface rotation than on the cases with solid-body surface rotation.

\begin{figure*}[!t]
  \centering

  $u_\phi$ to $u_{\left(r, \theta\right)}$ ratio \\
  \includegraphics[clip,trim=150 120 240 218, height=0.255\linewidth]{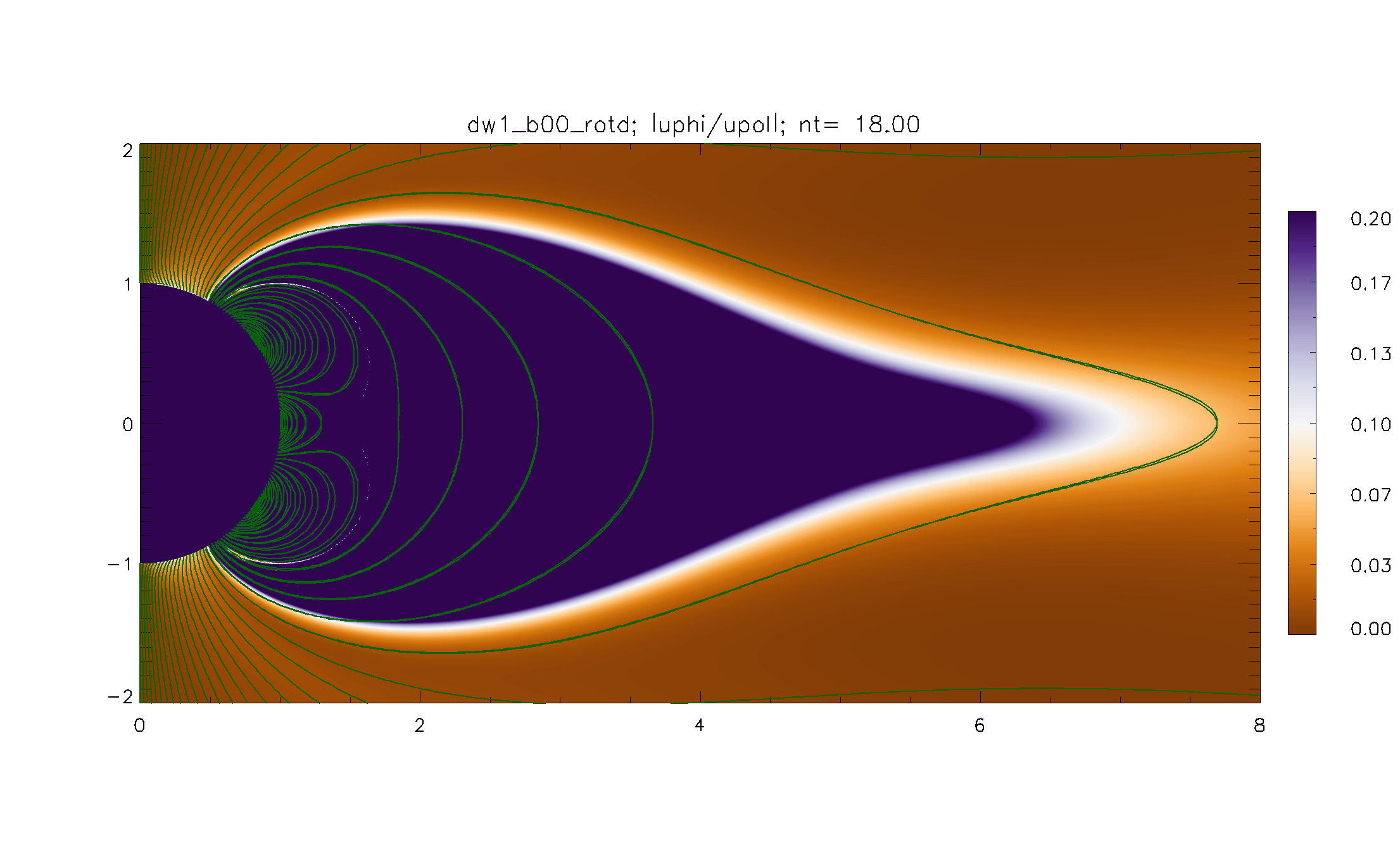}
  \includegraphics[clip,trim=150 120 5   218, height=0.255\linewidth]{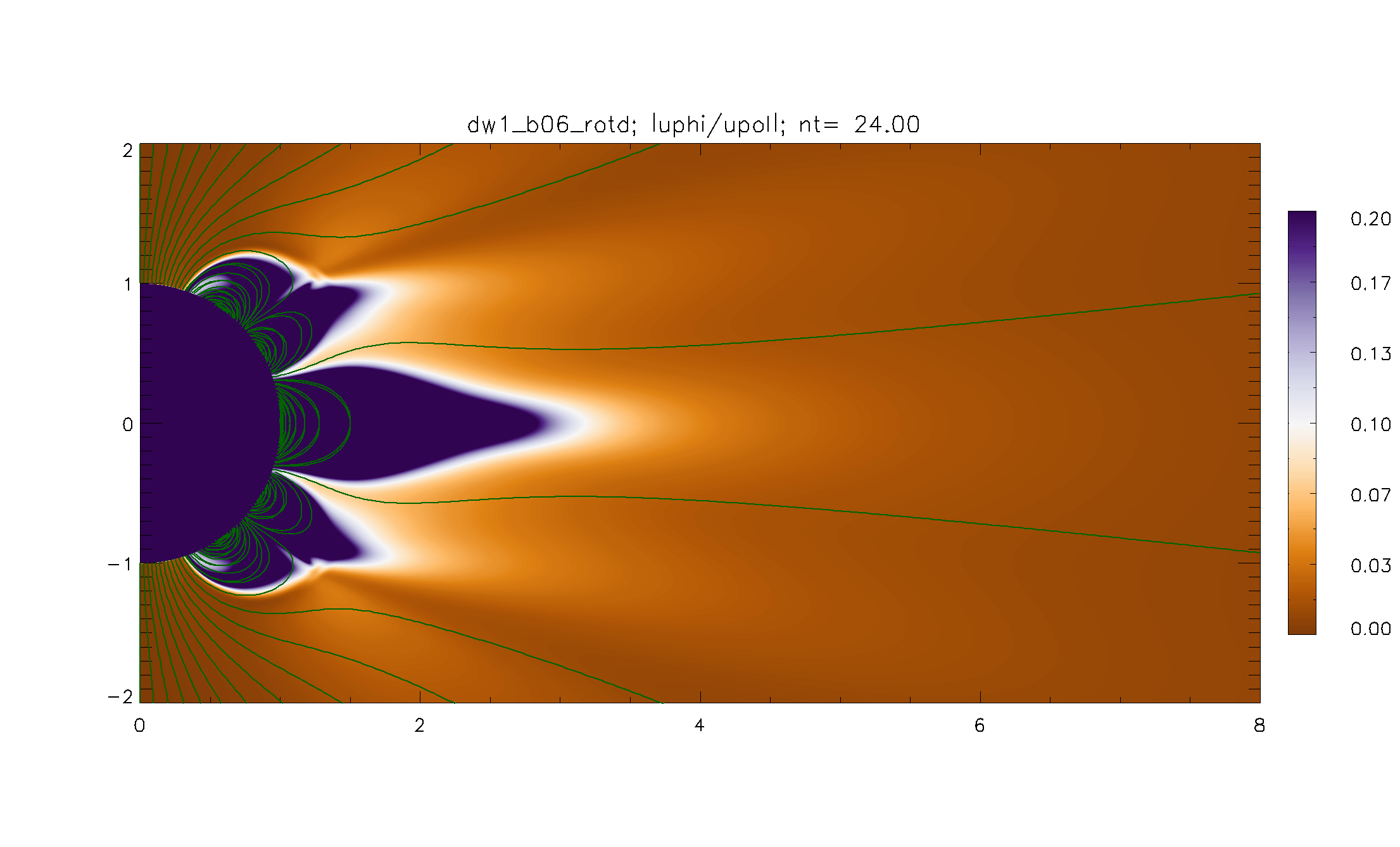}

  $B_\phi$ to $B_{\left(r, \theta\right)}$ ratio \\
  \includegraphics[clip,trim=150 120 240 218, height=0.255\linewidth]{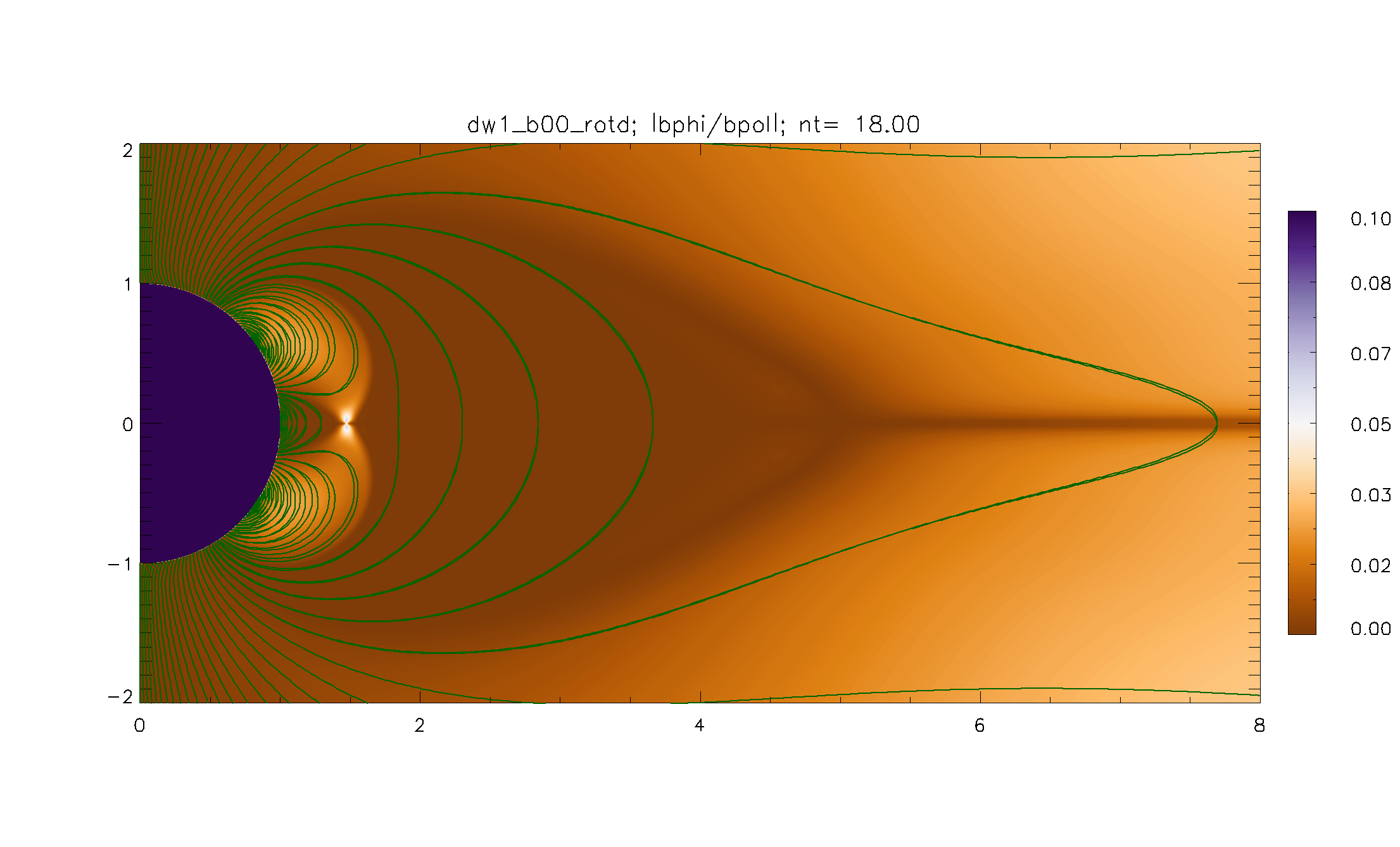}
  \includegraphics[clip,trim=150 120 5   218, height=0.255\linewidth]{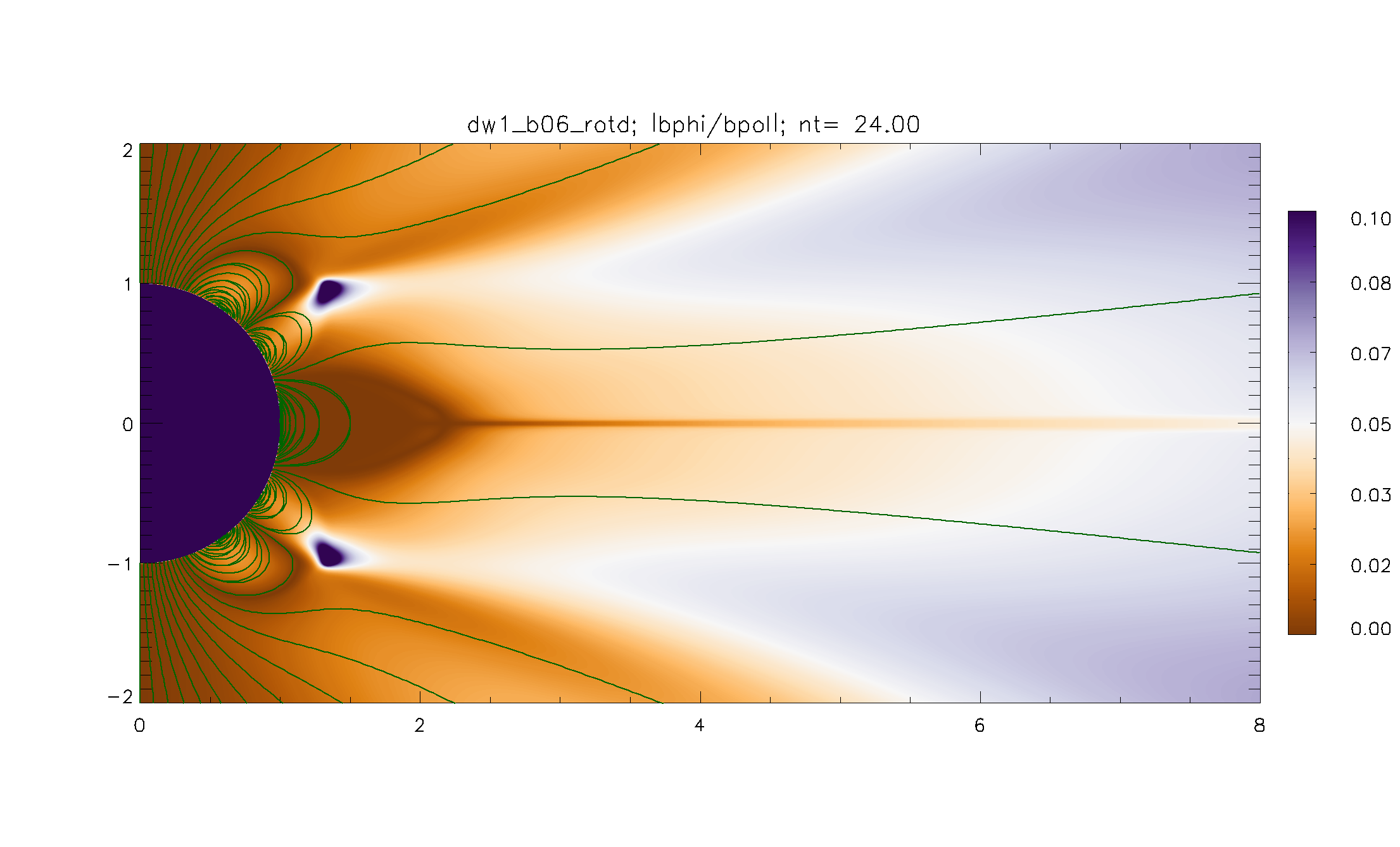}

  \caption{
    Ratio of azimuthal to meridional vector components of the velocity (top) and magnetic field (bottom) for the solar minimum (left) and solar maximum (right) configurations.
    Distances are in solar radii, and green lines are magnetic field-lines.
    For simplicity, only differentially-rotating cases are shown.
    The core of closed-field regions display high $u_\phi / u_{\left(r, \theta\right)}$ ratios and $B_\phi / B_{\left(r, \theta\right)} \approx 0$ due to them being in solid-body rotation.
    Open-field regions develop a $u_\phi / u_{\left(r, \theta\right)}$ profile that decays with altitude while $B_\phi / B_{\left(r, \theta\right)}$ grows.
    Coronal hole/streamer boundaries show stark contrasts in velocity and magnetic field pitch angles, and also extend coronal regions above the mid-latitude pseudo-streamers on the right panels.
  }
  \label{fig:vector_angles}
\end{figure*}

\subsection{Shear flow morphology}
In order to better show the form and amplitude of the shearing flows imposed by the global coronal rotation, Figure \ref{fig:flowlines_main} displays a series of velocity streamlines (or flowlines) corresponding to flows passing above and below the CH/streamer interface (from streamer mid-height to top).
The plots to the left show, for comparison, a non-rotating solar wind solution, while the plots to the right show the solar minimum configuration with solar-like differential rotation.
Three perspectives are presented: a front view, a side view, and a pole-on view (from the north pole), all in the inertial (not rotating) reference frame.
All streamlines are integrated for the same physical time interval, such that the length of each streamline indicates the total displacement of a given fluid element during that period of time.
Streamlines that correspond to plasma lying well within the streamer make an arc of a circle around the Sun (CCW on the pole-on view), while those connected to solar wind streams extend outward.
Among the latter, those that are farther away from the CH/streamer boundary cross the domain faster and follow a straighter path, while those that pass closer to it also accelerate slower and suffer a larger azimuthal deviation before they join the bulk of the wind flow above.
Some flows transition between the two regions, especially those passing very close to the interface for which diffusive processes favour that transition.
The side-view (second row) also shows how coronal rotation makes the equatorial streamer slightly taller and with a sharper transition to the neighboring coronal holes, on the region that develops the strongest magnetic and flow shear.

\subsection{Wind shear near the coronal hole boundaries and on the extended corona}

\begin{figure*}[!t]
  \centering

  Flow vorticity $\left|\nabla\times \mathbf{u}\right|$ \\
  \includegraphics[clip,trim=80 120 5 218, width=0.49\linewidth]{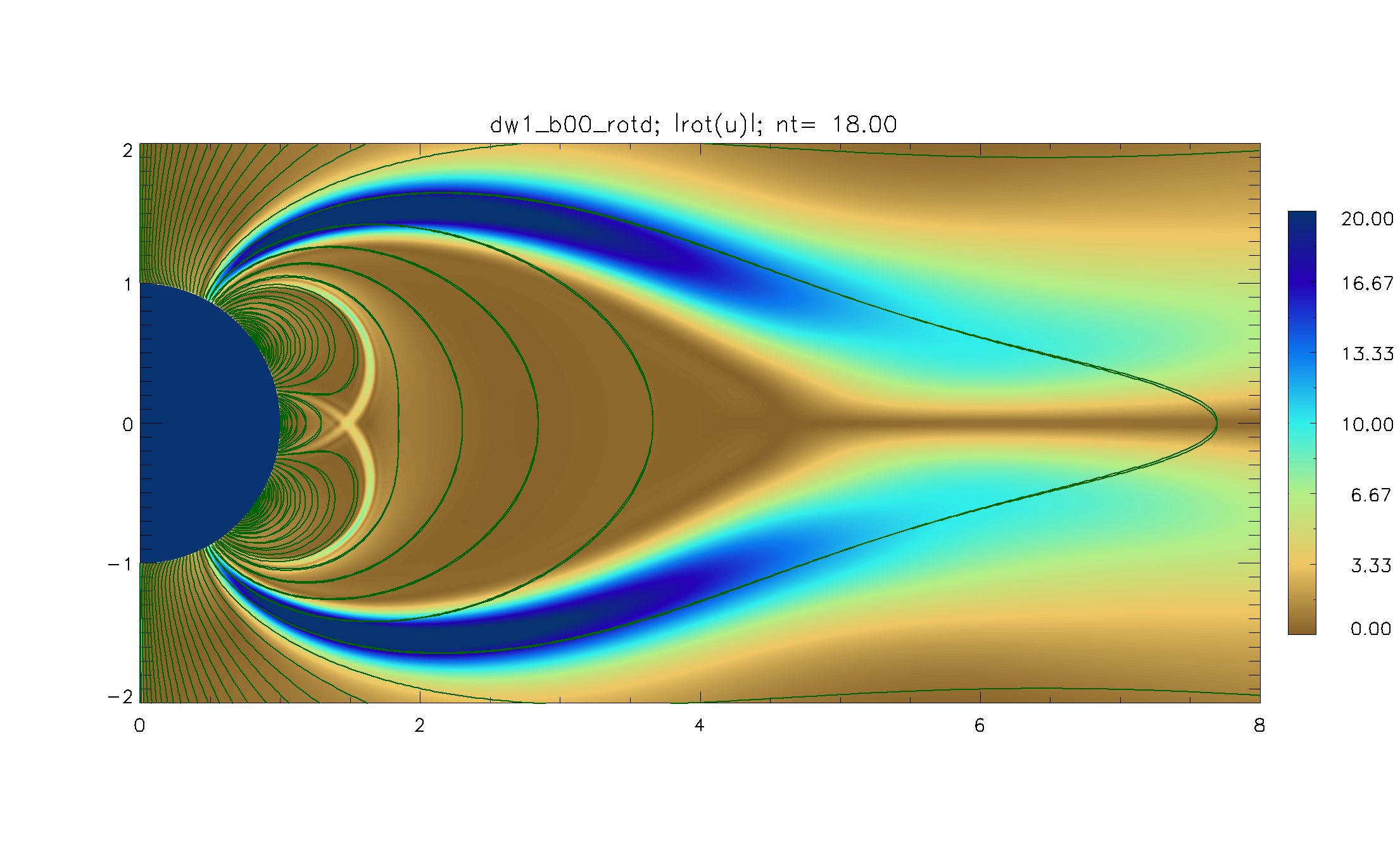}
  \includegraphics[clip,trim=80 120 5 218, width=0.49\linewidth]{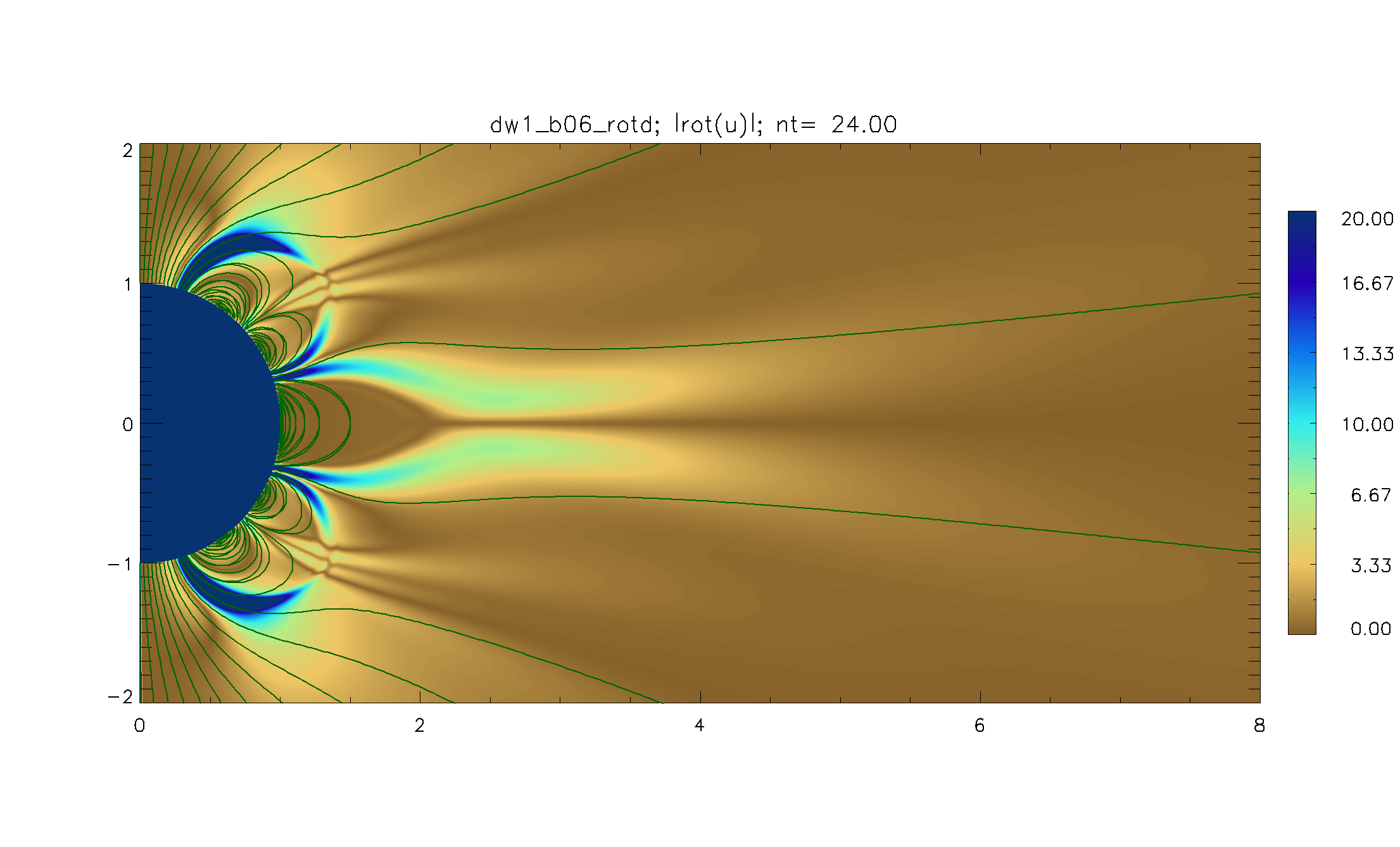}

  Ratio of poloidal (rotation-induced) to azimuthal (parallel flow induced) vorcity \\ 
  \includegraphics[clip,trim=80 120 5 218, width=0.49\linewidth]{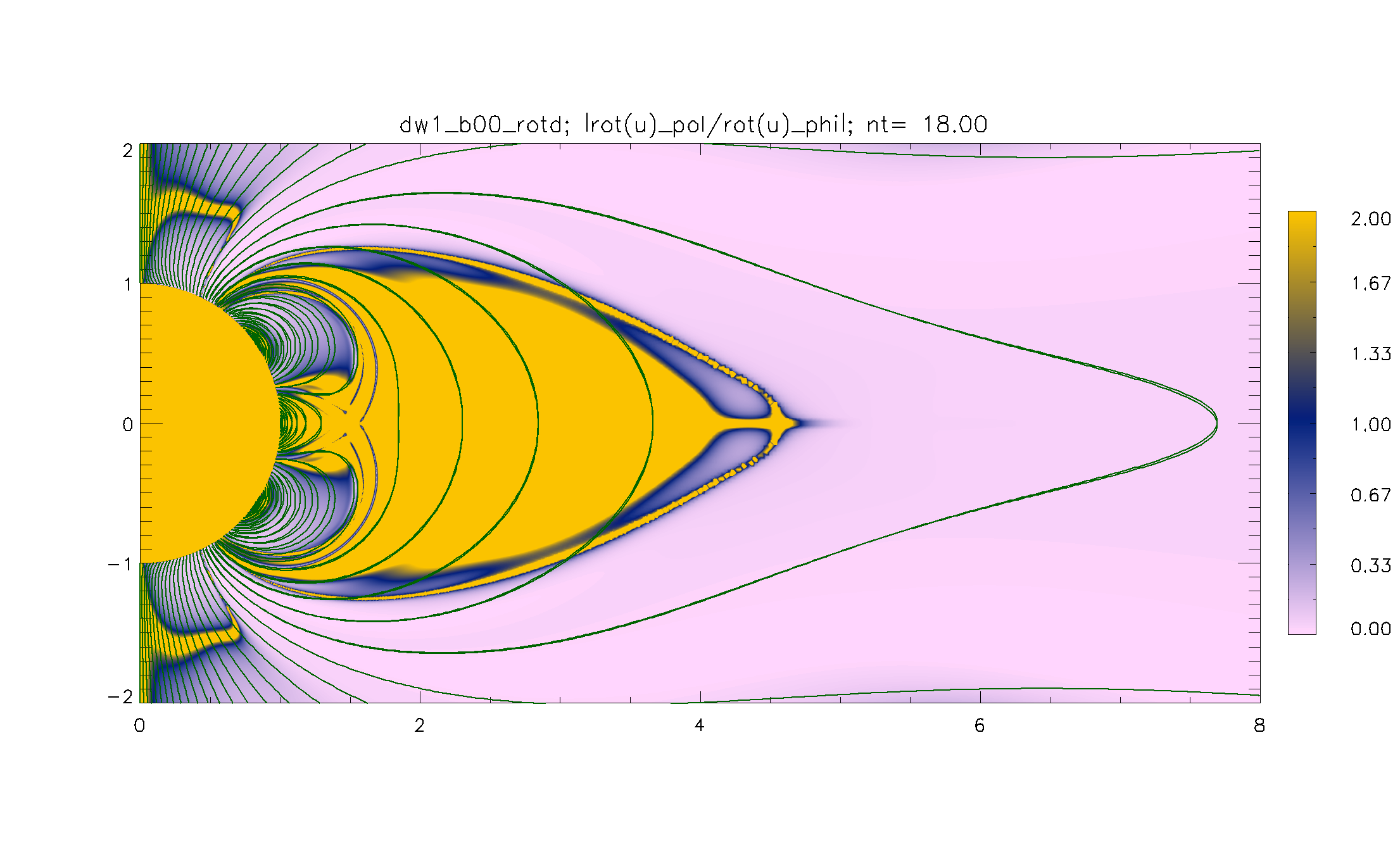}
  \includegraphics[clip,trim=80 120 5 218, width=0.49\linewidth]{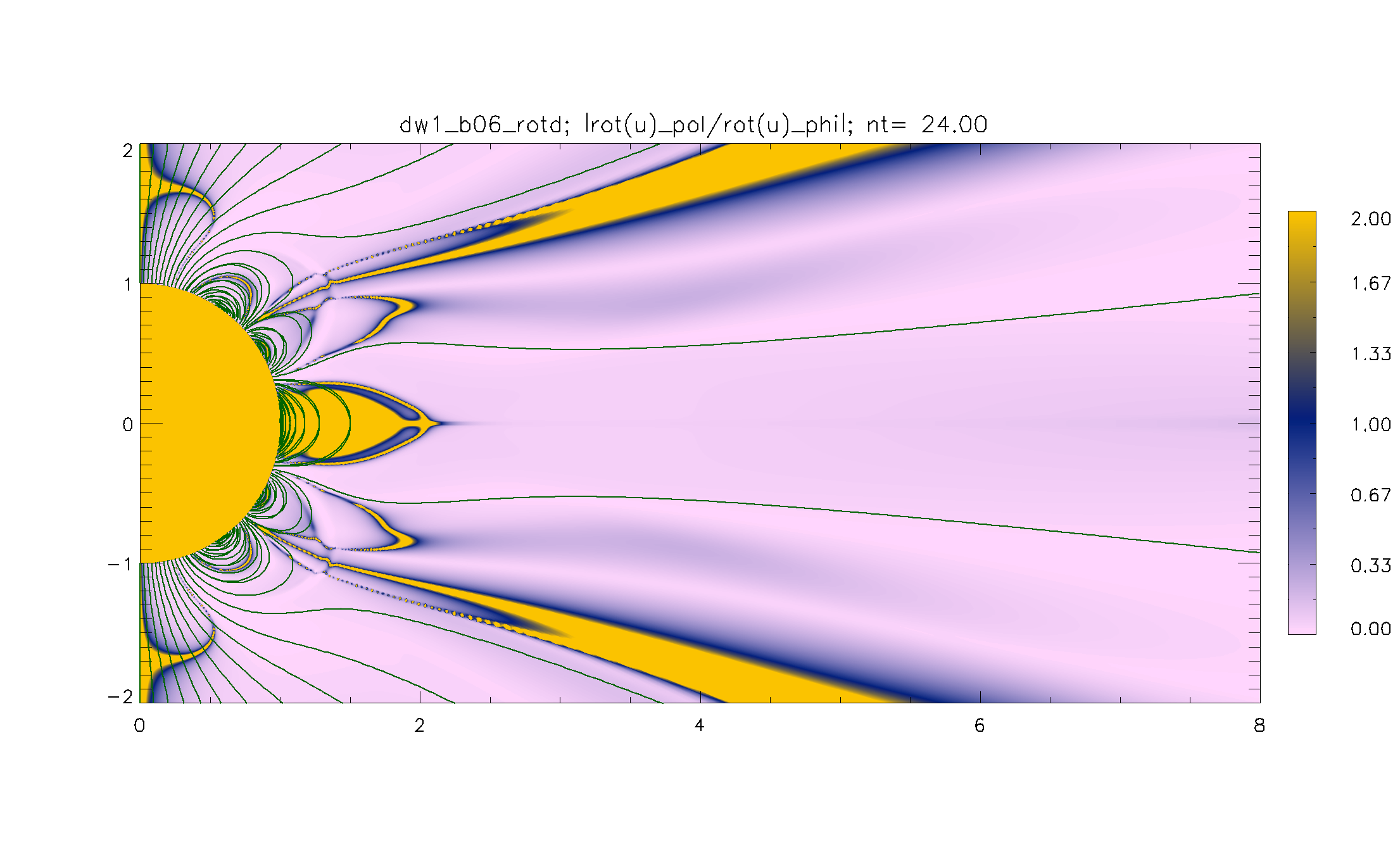}

    \caption{
      Unsigned flow vorticity (top) and ratio of poloidal to azimuthal vorticity components (bottom) in the \rev{low-latitude} regions for the solar minimum and solar maximum configurations.
    For simplicity, only differentially-rotating cases are shown and flow vorticity is displayed in normalised units ($9.0\e{-13}\un{rad\ s^{-1} m^{-1}}$).
    The axes indicates distances in solar radii. 
    Green lines are magnetic field-lines.
    The combination of the shearing flows in Fig. \ref{fig:flow_grad} are a source of vorticity that extend along the CH/steamer boundaries and above.
    The yellow regions in the bottom plots indicate zones where the poloidal component (the only one that gives rise to field-aligned vorticity, due to gradients of $v_\phi$ produced by coronal rotation) is predominant.
    Pink areas are dominated by wind speed shear (rather than rotational shear). 
    Streamers have a strong rotational signature.
    Pseudo-streamer axis display elongated rotational shearing zones.
  }
  \label{fig:vorticity_plots}
\end{figure*}

The CH/streamer interface region combines two mechanisms that generate shearing flows: one that acts on the meridional plane (due to the transition from the inner streamer stagnant flow, to the slow wind region and finally to the fast wind), and another one that translates into changes in the azimuthal component of the flow velocity (due to the different rotation regimes on each side of the boundary).
Figure \ref{fig:flow_grad} shows the absolute values of the gradients of the wind speed (top) and $\Omega$ (bottom) in the low corona during solar minimum (left) and solar maximum (right), focusing on the low latitude regions.
%
In absolute terms, flow shear is maximum on broad regions around the CH/streamer interfaces, that extend partially into the close-field regions and into the coronal holes. These broad regions of strong shear clearly extend upwards, surrounding the heliospheric currents sheets (HCSs).
The closed-field side of the shear region encloses the transition from the solidly-rotating but windless part of the domain to the slow wind zone, while the open-field side covers the transition from slow to fast solar wind flow (with decreasing rotation rates).
These shearing layers are longer and thicker on the solar minimum configuration, \rev{which} comprises a large equatorial streamer.
%
%
The contribution of coronal rotation to the overall shear is more clearly visible on the bottom row of the figure.
Strong gradients of $\Omega$ are found to be more tightly concentrated around the interface layers.
However, these gradients extend sideways well into the coronal holes, and upwards asides HCSs and also along pseudo-streamer axis (cf. the two mid-latitude pseudo-streamer structures on the panel to the right of the figure).
In addition, current density ($\sim \nabla\times\mathbf{B}$) also accumulates at the coronal hole/streamer boundaries, at the streamer tip and along the HCS.
Plasma $\beta$ increases with height along the interface, eventually becoming larger than 1 close to the streamer tip (and base of the HCS).

Angular deviations of magnetic field lines relative to the meridional plane (due to rotation) are small at these heights, and therefore hard to visualize directly.
But they are present nonetheless, and some of the plotted fieldlines can be seen to go in and out of the meridional planes represented in Fig. \ref{fig:mach_omega_panels}.
Open field lines bend backwards, as expected, and more strongly in the slow wind parts, while the closed loops on the other side of the interface are close to solid-body rotation.
From the high corona upward, the transverse (azimuthal) component of the magnetic field grows in amplitude with growing distance from the Sun in response to solar rotation, as expected from classic wind theory.
The azimuthal component of the velocity field, conversely, can be significant in the low corona, but decreases asymptotically in the high corona.
Fig. \ref{fig:vector_angles} shows the spatial distribution of the ratios between the azimuthal and the poloidal components of the velocity and of the magnetic field vectors ($u_\phi / u_{\left(r, \theta\right)}$ and $B_\phi / B_{\left(r, \theta\right)}$, in absolute value).
High values correspond to high pitch angles, defined as the angle between the vector and the meridional plane, or the angular deviation in respect to a purely poloidal flow (in $r$ and $\theta$).
A non-rotating corona displays null values everywhere on a non-rotating corona.
A rotating corona will contain closed-field region in (quasi) solid-body rotation with very high $u_\phi / u_{\left(r, \theta\right)}$ and $B_\phi / B_{\left(r, \theta\right)}\approx 0$,  and open-field regions for which $u_\phi / u_{\left(r, \theta\right)}$ decreases and $B_\phi / B_{\left(r, \theta\right)}$ increases with altitude.
Magnetic field-lines are represented as green lines.
Some magnetic features, such as the pseudo-streamers on the case represented on the right panels, generate localized enhancements of the transverse (azimuthal) fields that are felt across the radial extent of the numerical domain.

The combination of flow shears \rev{in the meridional and azimuthal} directions act as a persistent source of flow vorticity (defined as $\nabla\times \mathbf{u}$), also with multiple components.
Pure wind speed shear, due to variations in the radial and latitudinal components of the wind speed gradients, translates into an azimuthal vorticity vector that represents vortical motions contained within the $\left(r, \theta\right)$ plane.
On the other hand, spatial variations in the azimuthal speed (due to rotation) produce a vorticity vector with only radial and latitudinal components (poloidal vorticity), corresponding to vortical flows that develop in the $\left(\theta, \phi\right)$ or in the $\left(r, \phi\right)$ planes.
Figure \ref{fig:vorticity_plots} shows the spatial distributions of the absolute values of flow vorticity (top) and the ratio of the poloidal to azimuthal vorticity (i.e, ratio of rotation-induced to wind-speed induced vorticities; bottom plots).
Left and right columns show the differentially-rotating cases at solar minimum and at solar maximum, as in the previous figures.
Flow vorticity is maximal at the interfaces between closed and open magnetic field, as can be guessed from Figs. \ref{fig:flow_grad} and \ref{fig:vector_angles}, and extends outward around streamer and pseudo-streamer stalks.
The yellow regions in the bottom plots indicate zones where the poloidal component -- the one that is rotation-induced and can give rise to field-aligned vorticity -- is predominant.
The pink areas are, conversely, dominated by wind speed shear caused by the spatial distributions of slow and fast wind flows and of wind-less closed-field regions.
Streamers (and generally closed-field regions) have a strong rotational signature that is present up to their boundaries.
The accelerating wind flows that surround them rapidly reach speeds high enough to overcome the azimuthal (rotational) speeds, and most of the coronal hole vorticity becomes dominated by wind speed shear.
However, some coronal regions develop a strong and extended rotation-induced vorticity signature.
The most remarkable examples are pseudo-streamers \rev{located} at mid-latitudes, \rev{which} develop elongated rotational shearing zones along their magnetic axis, as shown in the bottom right panel of Fig. \ref{fig:vorticity_plots}.
These structures separate regions of the corona filled with magnetic field rooted at very different latitudes (polar and quasi-equatorial CHs) with rather different surface rotation rates.
Flow shear remains predominantly forced by global rotation through large distances on these layers, and thus developing a mainly field-aligned vorticity (corresponding to vortical motions orthogonal to the main magnetic field orientation).

\section{Sun -- Parker Solar Probe connectivity context}

\begin{figure*}[!t]
  \centering

  \includegraphics[clip,trim=1 1 15 0,width=\linewidth]{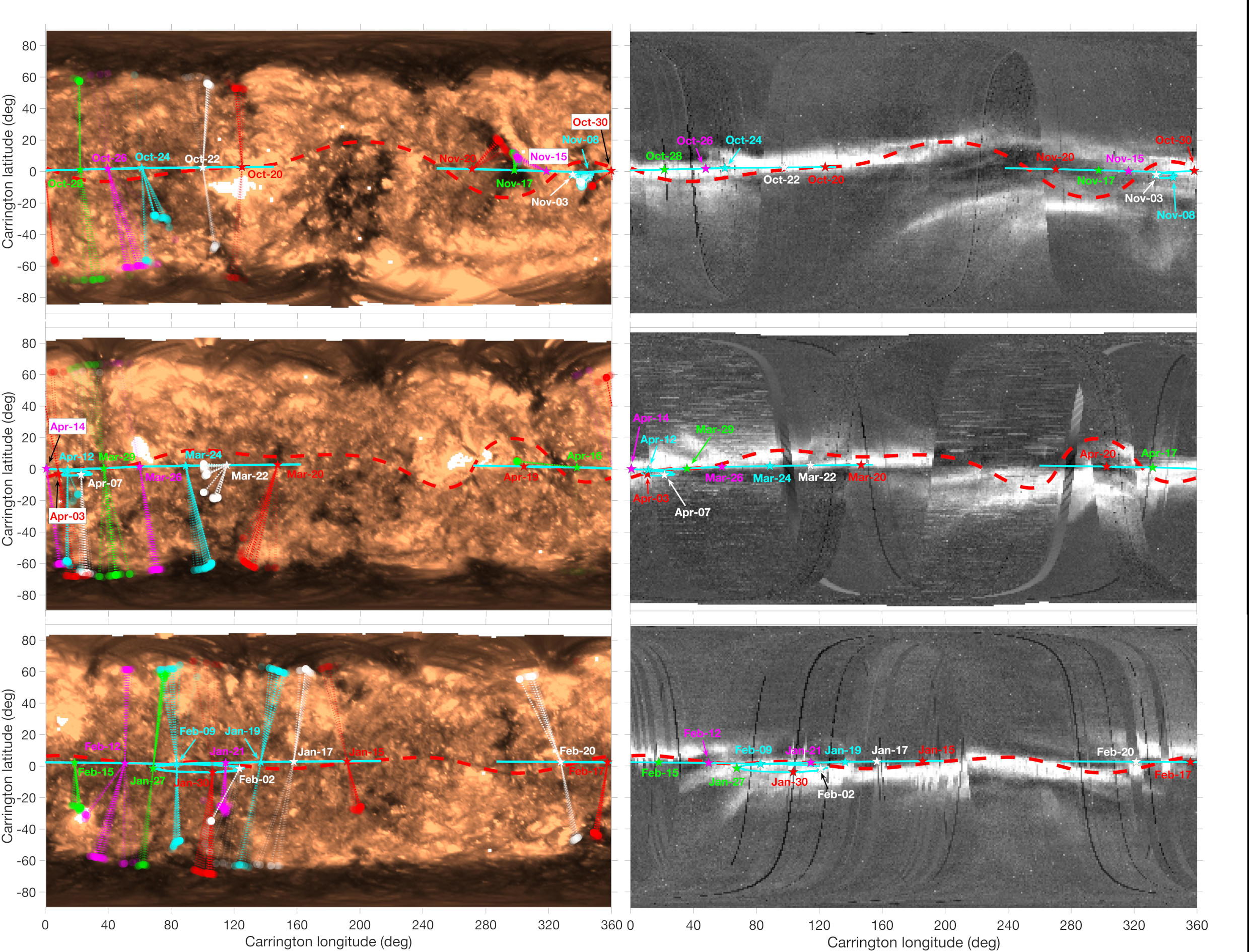}

    \caption{
      \rev{%
        Sun to PSP connectivity maps for the first (Oct. -- Nov. 2018), second (Mar. -- Apr. 2019) and fourth (Jan. -- Feb. 2020) encounters (top, middle and bottom rows, respectively).
        Connectivity is computed by projecting the solar wind speed measured at PSP backwards to the surface of the Sun,  considering the most likely Parker spiral, accounting for solar wind travel time and connecting to coronal field reconstructions based on magnetograms corresponding to the expected solar wind release times.
        The left panels display Carrington maps of EUV emission (SDO/AIA $193\AA$) overlaid with the most probable footpoints (solar wind source positions, coloured circles) at the surface of the Sun for the dates indicated, and the centroid of the connectivity probability distribution at $5\rsun$ (coloured stars).
        The cyan line indicates the trajectories of the $2.5\rsun$ connectivity point.
        The coloured dotted lines are visual aids to connect source-surface to surface positions (but do not trace the actual field-line trajectories).
        Date labels indicate in-situ solar wind measurement times.
        The right panels show similar maps of white light emission at $5\rsun$ (SoHO/LASCO C3), overlaid with similar markers (at $5\rsun$).
        The red dashed-line represents the neutral line (base of the HCS) obtained from PFSS extrapolation of ADAPT/GONG maps.
        PSP spent a significant fraction of its first passages sampling solar wind streams that developed at the vicinity of predominantly azimuthaly-aligned CH-streamer boundaries, and that propagated through the heliospheric plasma sheet (in close proximity to the HCS, within the bright streamer belt region).}
  }
  \label{fig:psp_connect_path}
\end{figure*}

\begin{figure*}[!t]
  \centering

  \includegraphics[width=0.49\linewidth]{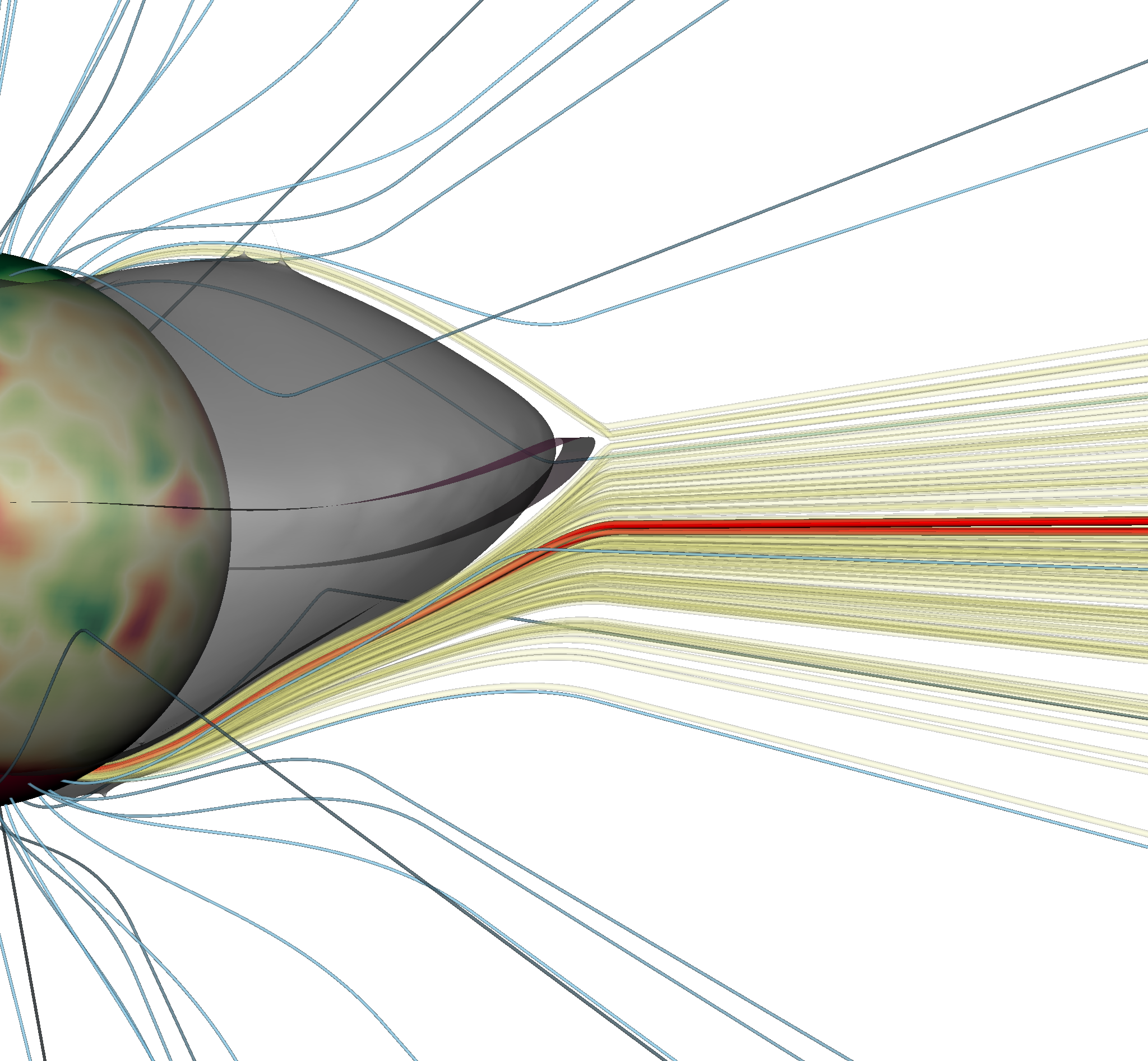}
  \includegraphics[width=0.49\linewidth]{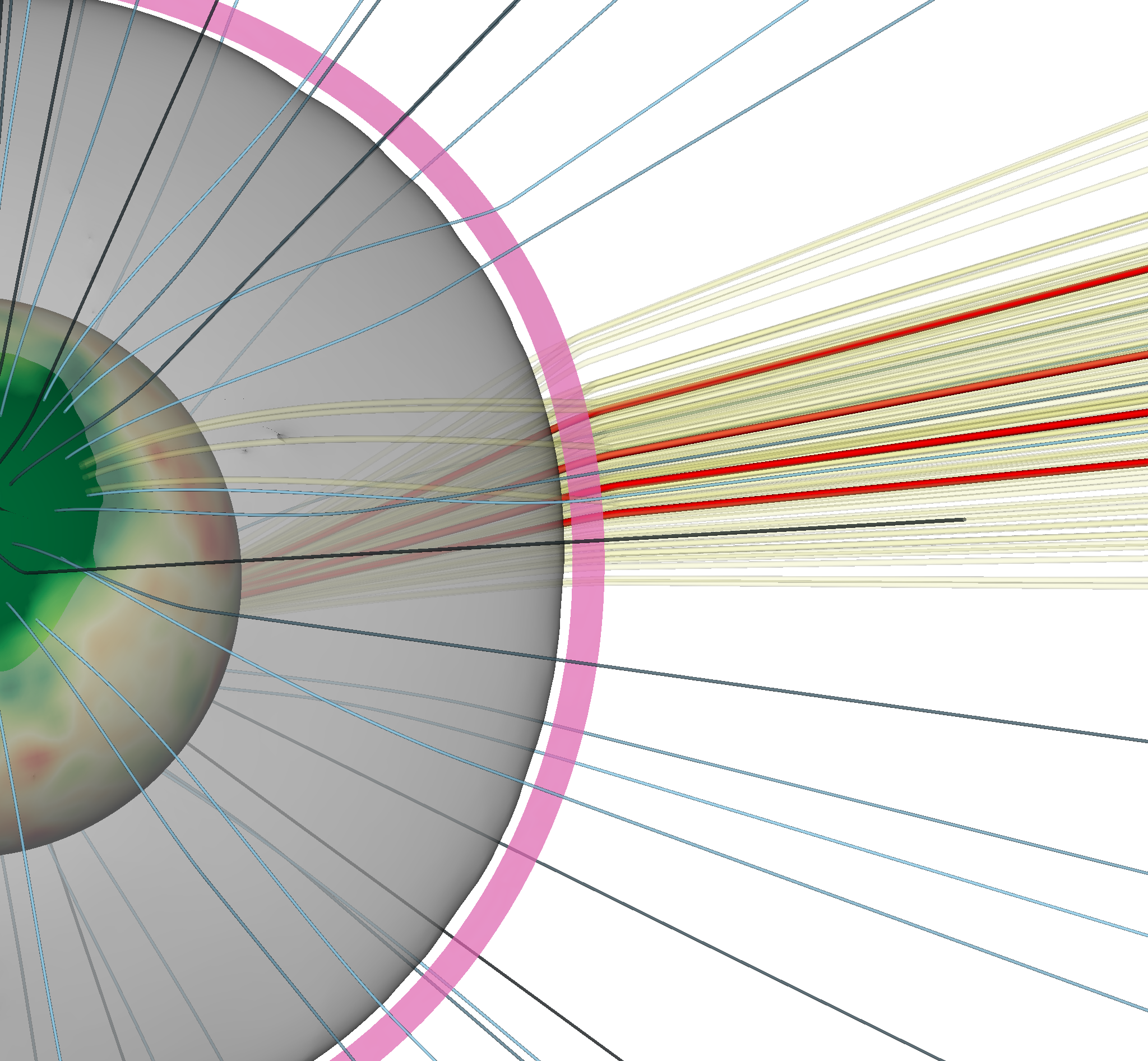}
  
  \caption{
    \rev{Three-dimensional rendering of the Sun-to-PSP connectivity paths on Mar 23th 2019 at 00:00 UT, during PSP's second encounter (\emph{cf.} middle panels of Fig. \ref{fig:psp_connect_path}).
      The two panels show side and top views, as for the simulations in Fig. \ref{fig:flowlines_main} (middle and bottom panels).
      The radial component of the surface magnetic field is represented in green (positive polarity) and red (negative polarity) tones, retrieved from the same ADAPT/GONG magnetograms as in the previous figure.
      The transparent grey surface represents the boundary between the large equatorial streamer and the polar coronal holes, and the violet ribbon indicates the base of the HCS.
      The light yellow lines cover the whole uncertainty ellipse taken into account by the Connectivity Tool.
      A fraction of these lines map towards the northern hemisphere, but can be discarded because they correspond to a magnetic polarity inverse of what PSP measured in-situ (although they fall into topologically and dynamically equivalent regions of the corona).
      The dark red lines indicate the connectivity paths ranked with the highest connectivity probability.
      The blue lines trace a few additional open field lines rooted inside coronal holes.
      PSP is connected to field lines (and wind streams) that closely delineate azimuthaly-aligned CH-streamer boundaries, in concordance with the MHD simulations discussed in Sect. \ref{sec:dip}.
      Note how the lines (than sample the whole uncertainty ellipse) clearly fall into compact regions that correspond to the broad shearing zones in the top-left panels of Figs. \ref{fig:flow_grad} and \ref{fig:vorticity_plots}.}
  }
  \label{fig:connectivity}
\end{figure*}

\rev{
  In order to establish links between the results of our numerical simulations and measurements made by Parer Solar Probe, we attempted to determine the source regions of the wind streams detected in-situ, together with their trajectories across the solar corona.
  This let us verify the applicability of the model to the coronal context at play, and hence to determine the physical conditions most likely experienced by those solar wind flows.
}

\subsection{The Connectivity Tool}
\label{sec:connecttool}

\rev{We have used the IRAP's Connectivity Tool} (\url{http://connect-tool.irap.omp.eu/}) to determine the most probable paths that the solar wind streams took all the way from the solar surface to the position of PSP during its first four perihelia.
The connectivity tool calculates continuously the magnetic and solar wind connectivity between the Sun and different spacecraft (or Earth), so as to establish physical links between them all along their orbits \citep{rouillard_models_2020}.
The tool offers different possible sources of input magnetogram data, extrapolation methods and wind propagation models, and lets the user assess uncertainties and inconsistencies related to the measurements and models used.
We chose to setup the connectivity tool with standard Potential Field Source-Surface (PFSS) extrapolations of Air Force Data Assimilative Photospheric Flux Transport (ADAPT) magnetograms \citep{arge_improvement_2000,arge_improved_2003}. It also provides an evaluation of the Potential Field Source-Surface (PFSS) reconstructions by comparing the topology of the neutral line with that of streamers (Poirier et al. 2021, \emph{submitted}). The choice of the ADAPT magnetograms was based on this evaluation procedure.
\rev{Propagation paths and temporal delays were determined by adjusting a solar wind profile to the wind speeds measured at the position of PSP at each moment of its trajectory. The wind velocities were obtained from the SWEAP instrument suite \citep{kasper_solar_2016}, and particularly from the plasma moments from the Solar Probe Cup \citep[SPC;][]{case_solar_2020}.}
\rev{%
  The solar wind mapping from the spacecraft position to the low corona, and from there to the surface, can be affected by different sources of error.
  Uncertainties related to the exact wind acceleration profile can lead to different solar wind propagation paths, and hence to deviations in longitude in the high corona, and total wind travel time.
  PFSS extrapolations from magnetic fields measured at the surface of the Sun are furthermore occasionally affected by positional errors that translate into few-degree deviations of the coronal structures in latitude.
  In order to cope with these issues, the Connectivity Tool determines the points at the surface of the Sun that connect to an uncertainty ellipse around the orbital position of the target spacecraft (covering the expected latitudinal and longitudinal uncertainties).
  For any given time, the tool hence provides a list of surface foot-points with different associated probabilities, rather than unique positions.
  The time period covered by this manuscript corresponds to solar minimum configuration, with a close to axi-symmetric corona and a rather flat HCS. 
  As a result, the dispersion of the foot point positions predicted for any given date and time leads to region very elongated in the azimuthal direction, with the corresponding errors rarely leading to topologically different regions of the Sun.
  This holds also for moments when the spacecraft is in close proximity to the HCS and the expected footpoints split into both solar hemispheres (northern and southern end-regions are topologically similar).
  We also made extra runs with lower and higher solar wind speeds for the duration of the time-periods analysed, and found that the properties of the connected regions do not change significantly.
  Our aim here is to identify the types of solar wind source regions (in the topological sense) rather than exact surface footpoint coordinates.
  Therefore the error sources described above do not translate into variations of the dynamical and topological properties of the regions crossed by the solar wind, and therefore should not affect our analysis.
}

\subsection{Sun to spacecraft connectivity, solar wind trajectories}
Figure \ref{fig:psp_connect_path} shows the trajectory of PSP during perihelia 1,2 and 4 (thick black line on the left, green on the right panels) plotted over Carrington maps of coronal EUV emission (SDO/AIA 193~\AA; first column) and white-light emission at $5\un{\rsun}$ (SoHO/LASCO C3; second column).
The corresponding time periods are during Oct. -- Nov. 2018 (top), Mar. -- Apr. 2019 (middle) and Jan. -- Feb. 2020 (bottom).
Pale blue and red layers cover the base of coronal holes with positive and negative polarity, respectively.
The colored symbols represent the positions of the solar wind plasma that reached PSP computed using the IRAP's connectivity tool, with in-situ measurement dates being labelled in corresponding colours.
Stars indicate solar wind plasma position at the PFSS source-surface altitude \rev{(only the centroid of the uncertainty ellipse is shown, for simplicity)}, and circles indicate the most likely wind source positions at the solar surface.
The dotted lines that connect them are visual aids and, for simplicity, do not trace the actual field-line trajectories.
\rev{Following the description in Sect. \ref{sec:connecttool}, each star corresponds to several circles that indicate the scatter due to mapping uncertainties.
  The red dashed-line represents the neutral line (base of the HCS) obtained from PFSS extrapolation of ADAPT/GONG maps.}
\rev{We have excluded the third PSP encounter due to the unavailability of continued good quality solar wind data during the corresponding time interval.}

\rev{The surface footpoint charts overlaid on the EUV 193~\AA\ synoptic maps indicate that PSP spent a significant fraction of its first passages sampling solar wind streams that developed at the vicinity of predominantly azimuthaly-aligned CH-streamer boundaries.
  These regions correspond topologically to the boundary shear layers illustrated in the top-left panel of Fig. \ref{fig:flow_grad} (for our simulated solar minimum case).
  The mappings at $5\rsun$ on the right panels (white-light maps) show that the streams detected by PSP propagated through the heliospheric plasma sheet (in close proximity to the HCS), and through the bright streamer-belt region.
These region corresponds to the equatorial regions in the high coronain our solar minimum simulations, just above the tip of the large equatorial streamer.}

\rev{
Solar wind connectivity across this range of heliocentric distances ($1$ to $5\rsun$) is better shown in Figure \ref{fig:connectivity} for one selected instant (during the second PSP encounter).
The figure shows a three-dimensional rendering of the global magnetic field structure of the corona on March 23rd 2019 (\emph{cf.} Fig. \ref{fig:psp_connect_path}) and the magnetic field lines through which the wind plasma that reached PSP at the instant represented is likely to have escaped from beforehand.
The two panels show side and top views, as in Fig. \ref{fig:flowlines_main} (middle and bottom panels) for our solar minimum MHD simulations.
The ADAPT/GONG magnetogram used on our coronal field reconstruction is plotted over the surface of the Sun, with green and red shades representing positive and negative polarities.
The transparent grey surface represents the boundary between the large equatorial streamer and the polar coronal holes, and the violet ribbon indicates the base of the HCS (polarity inversion line).
Light yellow lines represent a series of magnetic field lines that sample the whole uncertainty ellipse taken into account by the Connectivity Tool.
A fraction of these lines map towards base of the coronal hole on the northern hemisphere, but are discarded because they correspond to a magnetic polarity inverse of that measured in-situ by PSP.
The dark red lines indicate the connectivity paths ranked with the highest connectivity probability.
The blue lines trace a few additional open field lines rooted inside coronal holes.
It is clear from the figure that the wind streams sampled by PSP at this instant developed along paths that closely delineate azimuthaly-aligned CH-streamer boundaries, in concordance with the MHD simulations discussed in Sect. \ref{sec:dip}.
In view of the temporal sequence of Sun-to-PSP solar wind connectivity displayed in Fig. \ref{fig:psp_connect_path}, this configuration was the most common one throughout the periods analysed.}

%
\rev{The most probable Sun--spacecraft solar wind propagation paths lie, for the most, right along the boundary between the large equatorial streamer and the polar coronal holes.
  As PSP proceeded on its orbit, the source regions at the surface scanned this boundary continuously, unless during brief periods of connection to low latitude coronal holes or to deep equatorward polar CH extensions.
  PSP spent a large fraction of its first few encounters probing solar wind streams that formed in the vicinity of azimuthally aligned CH/streamer boundaries, and especially so during the 2nd and 4th encounters, with occasional connections to small low latitude coronal holes or equatorward polar coronal hole extensions (Griton, et al, 2020; \textsl{in press}).}
\rev{%
  The solar corona retained a high degree of axial symmetry throughout the time periods involved in this analysis.
  The HCS (and HPS) maintained a predominant E-W orientation in the solar regions connected to PSP.
  These reasons justify the use of the MHD simulations represented in Figs. \ref{fig:mach_omega_panels} to \ref{fig:vorticity_plots}.
  This is especially true for the solar minimum case (with an axi-symmetric large equatorial streamer).
  Deviations to this configuration, such as the coronal hole extensions and equatorial coronal holes visible near Carrington longitudes $280$ and $80$ in Fig. \ref{fig:psp_connect_path}, are in principle associated with pseudo-streamers such as those in our simulations for the cycle decay phase (right panels of Figs. \ref{fig:flow_grad} to \ref{fig:vorticity_plots}, at mid-latitudes), although perhaps with different sizes and orientations.
}
\rev{%
  We consider that the regions rooted at about $\pm 60 \un{deg}$ in our simulations for the solar minimum case (\emph{cf.} Figs. \ref{fig:flow_grad} to \ref{fig:vorticity_plots}, left) are the most representative of the source regions of the wind flows detected by PSP.
  These are coronal hole boundary regions oriented in the east-west direction (parallel to the direction of solar rotation).
  Solar wind streams originating from these regions are accelerated through an environment with significant and spatially extended solar wind speed and rotation shear, that also correspond to the peaks in rotation period in the left panels of Fig. \ref{fig:rot_periods_plots}.
}

\rev{The dynamical properties of these zones of the corona should have an impact on the properties of the wind measured \emph{in-situ}, if not being being responsible for some of their characteristics.}
On PSP data, transitions from streamer (i.e, boundary layer) to non-streamer (core of coronal hole) wind flows were accompanied by a clear decrease in the variablity of the wind \citep{rouillard_relating_2020}, both in frequency and amplitude of magnetic SBs and on the occurrence of strong density fluctuations.
\rev{This is suggestive that the physical conditions associated with coronal hole boundaries are favourable to the development of such perturbations.
Interchange reconnection, often invoked as a possible SB generation mechanism \citep[][]{fisk_global_2020}, relies on the forcing of these boundary regions by the large-scale rotation of the corona.
However, it is expected to be enhanced (or more efficiently driven) at CH/streamer boundaries that are orthogonal or inclined with respect to the direction of rotation as, e.g, a streamer pushes into neighbouring CHs \citep[see e.g.][]{lionello_effects_2005}, and reduced on azimuthally aligned CH/streamer boundaries.
This is in contrast with the general spatial orientation of the equatorial streamer observed during this period and that of our simulations, on which streamer/CH boundaries are parallel to the direction of rotation.
While less favourable to interchange reconnection, this spatial orientation does not hamper the formation of the shear flows in the solar wind shown in Figs. \ref{fig:flowlines_main} to \ref{fig:vorticity_plots}.
Furthermore, these shears remain visible up to large heliocentric distances, which could favour the propagation (or even amplification) of magnetic perturbations formed in the low corona, allowing them to survive more easily up to the altitude of detection \citep{owens_signatures_2020,macneil_evolution_2020}.
Beyond this phenomenology, the boundaries of polar coronal holes are also known to be highly dynamic and undergo significant reconfiguration over the roughly 24-hour timescale of supergranules \citep{wang_formation_2010}.
These effects are not modelled in this paper, but could induce additional variability that could be felt by PSP.
The existence of neighbouring solar wind streams with different rotation rates (as those in the streamer stalk regions on our simulations) should furthermore contribute to increasing the variability of the transverse (rotational) velocities measured by PSP, especially as the HCS is slightly warped (\emph{cf.} Fig. \ref{fig:psp_connect_path}).
}

\section{Conclusion}
\label{sec:conclusion}

\subsection{Summary}

We investigate the development of spatially extended solar wind shear regions induced by solar rotation and by variations in solar wind speed in the light of recent Parker Solar Probe findings.
Our analysis combines simulations made using a MHD numerical model of the solar wind and corona,  and estimations of the sun-to-spacecraft connectivity during the first 4 PSP solar encounters to aid in associating model results to s/c data.
Our main findings are that:
\begin{enumerate}
\item Solar wind flows that develop in the vicinity of coronal hole boundaries are subject to persistent and spatially extended shearing.
  There are two components to this shearing: a wind speed shear due to the transition from closed-field (no-wind) to the slow and fast wind regions, and a rotational shear due to the way coronal rotation settles in response to the rotation of the solar surface.
  The most significant shearing occurs in thin layers that lie along CH/streamer (or pseudo-streamer) boundaries, and that stretch outwards in the vicinity of heliospheric current sheets and pseudo-streamer stalks.
  Wind speed shear generates a spatially broader shearing signature associated with a large-scale vorticity vector oriented in the azimuthal direction
  Rotational shearing produces shearing patterns with field-aligned vorticity that become predominant in elongated regions above pseudo-streamer stalks that extend to great distances from the Sun.    \\

\item The solar corona acquires a complex rotation pattern that differs significantly from that of the surface rotation that drives it.
  Closed-field regions (streamers, pseudo-streamers) tend to set themselves into solid body rotation, with a rate consistent to that of the surface regions at which they are rooted.
  Open field lines show a variety of rotation rates, with those that pass near coronal hole boundaries  acquiring the lowest rotation rates at mid coronal heights, in stark contrast with the closed-field regions across the boundary. This results in clear increases of rotation period adjacent to streamers (especially visible in large streamers), in agreement with SoHO/UVCS observations.
  Streamer stalks (and the vicinity to HCS/HPS) can contain a mixture of wind streams at different rotation rates (slow rotating flows coming from the CH boundaries, faster wind flows coming from the streamer tips).\\

\item Solar wind flows probed by Parker Solar Probe during its first four orbits form and propagate away from the Sun through regions of enhanced wind speed and rotational shear.
  Our Sun-to-spacecraft \rev{connectivity analysis} shows that such solar wind flows originated mostly at the boundaries of quasi-axisymetric polar coronal holes, with occasional crossings of low-latitude coronal holes.
  The measured wind flows showed a strong and complex rotational signature permeated by pervasive magnetic perturbations such as switchbacks (among others).
  \rev{Our results suggest that the slow wind flows detected by PSP should be experience persistent shears across their formation and acceleration regions, supporting the idea that these should have an impact} on the formation localised magnetic field reversals and be favourable to their survival across the heliosphere.
\end{enumerate}

\subsection{Discussion and perspectives}

Unlike its photospheric counterpart, current knowledge of the rotational state of the solar corona is very limited.
Different observation campaigns (relying on different measurement methods) have suggested that the corona rotates in a manner that is not in direct correspondence with the surface rotation, and that it depends to some degree on the solar cycle phase (that is, on the global magnetic topology).
More recently, Parker Solar Probe unveiled the prevalence of strong rotational flows that increase rapidly in amplitude as the spacecraft approached the Sun, at least in the regions probed during its first few close passes \citep{kasper_alfvenic_2019}.
%
%
Our MHD simulations show that surface rotation is transmitted to the corona in a complex manner that depends intrinsically on the organisation of the large-scale magnetic field at any given moment.
Coronal rotation is highly structured at low coronal altitudes, with a clear signature of slowly rotating flows that follow the CH/streamer boundaries, in agreement with the observations by \citet{giordano_coronal_2008}.
Some of these strong gradients in rotation rate produce an imprint that extends faraway into the high corona (up to the upper boundary of the numerical model).
These non-uniformities in rotation rate translate into solar flow shear with a vorticity component oriented along the magnetic field and solar wind propagation direction, that adds up to the shear caused by the spatial distribution of fast and slow wind flows (and that can only generate an orthogonal vorticity component).

The MHD model setup \rev{relies upon} a number of simplifications to the full physical problem.
It \rev{uses} a polytropic description of the plasma thermodynamics, meaning that the heating and cooling mechanisms are not modelled in detail, but that the main dynamical and geometrical features of the solar wind are retained.
This approach furthermore leads to a solar wind with speed variations which are much weaker than those found on the real solar atmosphere (smaller contrast between typical fast and solar wind speeds and broader transitions).
We settled on the limiting case in which the rotating solar corona and solar wind are perfectly axisymetric (with CH/streamer boundaries perfectly parallel to the direction of rotation).
Rotation-induced interchange reconnection is completely inhibited in this configuration, as is the development of shear instabilities.
The formation of full vortical flows in the $\left(\theta, \phi\right)$ plane and the injection of helicity into the wind flow are inhibited.
This choice of problem symmetry \rev{has}, however, a number of advantages in respect to the full 3D equivalent, namely that it allows running many more variations of parameters (different phases of the cycle, different solar surface rotation profiles), and that the runs can be easily made at a higher spatial resolution.
\rev{As a} consequence, the simulations develop sharper CH/streamer boundaries \rev{than their full 3D counterparts}, which help making the rotation shears more apparent.
These boundaries should, notwithstanding, be much sharper in the real solar corona.
For these reasons altogether, the amplitudes of solar wind speed and rotational shearing layers should be higher in the real Sun than those that the MHD simulations are capable of producing.
We also used an idealised solar dynamo model to \rev{constrain} the large-scale magnetic field topology at each moment of the solar cycle, meaning that our MHD simulations do not aim at modelling a specific event, but rather at letting us understand the dynamics of the regions of interest (the model produces a full set of typical solar coronal structures -- streamers, pseudo-streamers and coronal holes -- placed at different latitudes according to solar activity).

As we have shown with the help of the IRAP's connectivity tool (\url{http://connect-tool.irap.omp.eu/}), the solar wind streams that reached Parker Solar Probe during its first few encounters most often traversed the vicinity of the boundaries between a large equatorial streamer and the adjacent coronal holes.
The dynamics at play in such regions suggest that they are potential hosts for the development of shearing instabilities (among others) that are continuously driven by the large-scale wind shears at these boundaries.
Such processes can potentially introduce alfvénic perturbations into the solar wind and/or give rise to discrete helical (and pressure-balanced) MHD structures (depending on the aforementioned instability thresholds, growth rates and time-scales of ejection onto the wind).
Due to the limitations intrinsic to the global MHD approach expressed above, the simulations do not let us directly identify the onset of shear instabilities (such as Kelvin-Helmhöltz).
The velocity and magnetic field gradients remain, in practice, much smoother than those that real solar wind can have, and the global flow remains rather laminar (without fully developed turbulence). This is issue will be addressed in future work.
Interestingly, however, the solar wind shear -- due both to the transition from slow to fast wind and to the rotation rate gradients -- extends well beyond the height of the highest streamers in our simulations, meaning that the driving mechanism could be effective over a rather large range of heights.
One significant feature revealed by our simulations is the formation of a very long column with a strong field-aligned vorticity signature atop of a pseudo-streamer (bottom right panel in Fig. \ref{fig:vorticity_plots}).
Coronal rotation forces a strong shear across the pseudo-streamer axis in the plane perpendicular both to the wind flow and magnetic field direction that acts on the solar wind over large distances.
This provides, perhaps, a more efficient mechanism to inject helicity (and/or perturbations in the transverse direction, such as those that characterize switchbacks) than the milder (and effective over shorter distances) shearing present at the boundaries of large equatorial streamers.
During its first solar encounter, Parker Solar Probe switched connectivity temporarily from an equatorial streamer boundary region to a low latitude pseudo-streamer that formed near the small equatorial coronal hole visible in the top left panel of Fig. \ref{fig:connectivity}.
During this period, PSP magnetic field measurements show a transition from a period of strong amplitude and very frequent switchbacks to another with lower amplitude and more spaced events, hence supporting the hypothesis above.
Our results therefore give support to a hybrid view on the origin of magnetic switchbacks, by providing both a generation mechanism acting in the low corona, and a way to sustain (if not amplify) at least part of them on their way through the high corona and heliosphere.

A more detailed investigation of these \rev{hypothetical} mechanisms is yet to be undertaken, and will require substantial modeling efforts.
We believe, however, that the scenario we describe in this manuscript can shed light on how to bring different characteristics of SBs (generation, propagation) come together harmoniously, and also provides new insights on the mechanisms related to plasma transport between closed and open field regions of the solar corona.
We look forward to Solar Orbiter \citep{muller_solar_2020} and to its combined in-situ and remote sensing campaigns\rev{, that will be made from increasingly higher latitudes and at different phases of the solar cycle, hence providing a unique opportunity to detect wind flows that are formed and accelerated in a larger range of coronal contexts.}
\rev{Parker Solar Probe will keep reducing its perihelium distance, and will certainly provide a closer look into the effects of solar wind shear and rotation on increasingly more pristine wind flows}.

\begin{acknowledgements}
We acknowledge support from the French space agency (Centre National des Études Spatiales; CNES;
\url{https://cnes.fr/fr}) that funds activity of the space weather team in Toulouse (Solar-Terrestrial Observations and Modelling Service; STORMS; \url{http://storms-service.irap.omp.eu/}) that developed the Connectivity Tool.
The work of L. G., A. P. R., and N. P. was funded by the ERC SLOW\_SOURCE project (SLOW\_SOURCE -- DLV-819189).
R. F. P. and A. S. B. acknowledge support by the ERC synergy grant Whole Sun \#810218.
The numerical simulations were performed using HPC resources from CALMIP (Grant 2020-P1504).
\end{acknowledgements}

\bibliographystyle{aa}
\bibliography{refs}



\end{document}